\documentclass[aps,prd,10pt,tightenlines,amsmath,unsortedaddress,twocolumn,amssymb, 10pt]{revtex4}
\usepackage{float}
\usepackage{graphicx}
\usepackage{dcolumn}
\usepackage{bm}
\usepackage{bbold}
\usepackage{enumerate}
\usepackage[utf8]{inputenc}
\usepackage[english]{babel}
\usepackage{amsthm}

\usepackage{xcolor}

\begin{document}

\title{Photon-ALP oscillations inducing modifications to photon polarization}

\author{Giorgio Galanti}
\email{gam.galanti@gmail.com}
\affiliation{INAF, Istituto di Astrofisica Spaziale e Fisica Cosmica di Milano, Via Alfonso Corti 12, I -- 20133 Milano, Italy}

\date{\today}

\begin{abstract}

Axion-like particles (ALPs) are very light, neutral, spin zero bosons predicted by many theories which try to extend and complete the standard model of elementary particles. ALPs interact primarily with two photons and can generate photon-ALP oscillations in the presence of an external magnetic field. They are attracting increasing interest since photon-ALP oscillations produce deep consequences in astrophysics particularly in the very-high-energy (VHE) band, where they increase the transparency of the Universe to VHE photons by partially preventing absorption caused by the extragalactic background light. Furthermore, ALPs explain why photons coming from flat spectrum radio quasars (a particular class of active galactic nuclei, AGN) have been observed for energies above $30 \, \rm GeV$ -- which represents a first hint for the existence of an ALP. In addition, ALPs solve an anomalous redshift dependence of blazar (an AGN class) spectra -- which represents a second hint for the existence of an ALP. In this paper, we study another effect of the photon-ALP interaction: the change of the polarization state of photons. In particular, we study the propagation of the photon-ALP beam, starting where photons are produced -- we consider photons generated in a galaxy cluster or in the jet of a blazar -- crossing several magnetized media (blazar jet, host galaxy, galaxy cluster, extragalactic space, Milky Way) up to its arrival at the Earth, where photons can be detected. In the presence of the photon-ALP interaction, we analyze the final photon survival probability $P_{\gamma \to \gamma}$ and the corresponding photon degree of linear polarization $\Pi_L$ for observed energies in the range $(1-10^{15}) \, \rm eV$ dividing it into three energy bands: (i) UV-X-ray band ($10^{-3}\, \rm {\rm keV}-10^2 \, \rm keV$), (ii) high-energy (HE) band ($10^{-1}\, \rm {\rm MeV}-10^4 \, \rm MeV$), (iii) VHE band ($10^{-2}\, \rm {\rm TeV}-10^3 \, \rm TeV$). We observe that those photons, which are expected as unpolarized in the absence of ALPs, are made partially polarized by the photon-ALP interaction, which generally modifies the initial photon degree of linear polarization $\Pi_{L,0}$ in a sizable and measurable way. Our findings can be tested by observatories like IXPE (already operative), and by the proposed missions eXTP, XL-Calibur, NGXP and XPP in the X-ray band and by the proposed missions COSI (approved to launch), e-ASTROGAM and AMEGO in the HE range. A possible detection of a departure of the photon polarization from the standard expectations would represent an additional {\it hint} for the existence of an ALP. We also discover a peculiar feature in the VHE band, where photons at energies above $ \sim (1-10) \, \rm TeV$ are fully polarized because of the photon-ALP interaction. A possible detection of this feature would represent a {\it proof} for the existence of an ALP, but, unfortunately, current technologies do not allow yet to detect photon polarization up to so high energies.

\end{abstract}

\keywords{axion; polarization}

\pacs{14.80.Mz, 13.88.+e, 95.30.Gv, 95.30.-k, 95.85.Pw, 95.85.Ry, 98.54.Cm, 98.65.Cw, 98.70.Vc}

\maketitle



\section{Introduction}

The understanding of the forces governing our Universe and of the nature of the constituent particles is at least incomplete: much evidence has been established for the existence of dark matter and dark energy as dominant elements of our Universe. Very promising candidates for dark matter~\cite{preskill,abbott,dine,arias2012} are represented by the so-called axion-like particles (ALPs, for a review see e.g.~\cite{alp1,alp2}), which are hypothetical very light particles predicted among many theories by the superstring theory~\cite{string1,string2,string3,string4,string5,axiverse,abk2010,cicoli2012}. ALPs are a generalization of the axion, the pseudo-Goldstone boson associated with the global Peccei-Quinn symmetry ${\rm U}(1)_{\rm PQ}$ which was proposed as a natural solution to the strong CP problem (see e.g.~\cite{axionrev1,axionrev2,axionrev3,axionrev4}).

ALPs differ from axions in two aspects: (i) while the mass and the coupling constant to photons are related quantities for the axion, the ALP mass $m_a$ and the two-photon-ALP coupling constant $g_{a \gamma \gamma}$ are {\it unrelated} parameters, (ii) while the axion necessarily couples to fermions and to gluons in order for the Peccei-Quinn mechanism to work, ALPs interact primarily with two photons -- other interactions are subdominant and can be safely discarded. Thus, photons traveling in a magnetized medium mix with ALPs through the coupling $g_{a \gamma \gamma}$ -- the magnetic field is necessary in order to compensate for the spin mismatch between photons and ALPs -- producing two different effects on the photon propagation: (i) photon-ALP oscillations~\cite{sikivie1983, raffeltstodolsky} similar to the oscillations of different flavor massive neutrinos, (ii) the change of the polarization state of photons~\cite{mpz, raffeltstodolsky}. Therefore, ALPs have a huge impact in astrophysics and especially in the very-high-energy (VHE) band: whenever the medium crossed by photons is filled by intense magnetic fields and/or the path inside a magnetized medium is long (for reviews, see~\cite{gRew, grRew}), photon propagation gets modified producing a long list of effects.

The photon-ALP conversion can occur inside different magnetic fields such as in that of the jet of an active galactic nucleus (AGN), where a sizable amount of ALPs can be produced~\cite{trg2015} explaining why flat spectrum radio quasars (FSRQs, a particular class of AGN) have been observed for energies above $30 \, \rm GeV$~\cite{trgb2012} -- which represents a {\it first} hint for the ALP existence. In addition, the photon-ALP conversion can occur in the turbulent magnetic field of a galaxy cluster resulting in irregularities in the observed spectra which have however not been detected yet~\cite{fermi2016,CTAfund}, inside the extragalactic space enhancing the transparency of the Universe for energies above $\sim 100 \, \rm GeV$~\cite{drm2007,dgr2011,grExt} partially preventing hard photon absorption caused by their interaction with the extragalactic background light (EBL) photons~\cite{franceschinirodighiero}, in the Milky Way or in all these media producing sizable alterations in the observed blazar (a type of AGN) spectra~\cite{gtre2019,gtl2020}. In addition, a {\it second} hint for the ALP existence comes from the solution of the problem of the anomalous redshift dependence of blazar spectra thanks to the introduction of the photon-ALP interaction~\cite{grdb}. Furthermore, the photon-ALP interaction produces consequences on stellar evolution~\cite{globclu}, it has been invoked to explain in galaxy clusters the spectral distortions of the continuum thermal emission ($T \sim 2\, \rm {\rm keV}-8 \, \rm keV$)~\cite{thermal} and the unexpected spectral line at $3.55 \, \rm keV$ as dark matter decay into ALPs and subsequent oscillations to photons~\cite{DMdecay}. ALPs have also been employed to describe a blazar line-like feature~\cite{wang}. If confirmed, the detection of the gamma-ray burst GRB 221009A at $18 \, \rm TeV$ by LHAASO~\cite{LHAASO} or even at $251 \, \rm TeV$ by Carpet-2~\cite{carpet} would represent a strong indication for the ALP existence with the properties of the previous two hints~\cite{grtGRB}.

All the above-listed ALP effects are linked in some way to the modification of the amount of observable photons because of the photon-ALP oscillations but also the second main effect of the photon-ALP interaction i.e. the change of the polarization state of photons possesses strong implications for a possible ALP indirect detection. In fact, whenever the polarization of the detected photons differs with respect to conventional physics expectations, this fact may represent a hint for new physics in the form of ALPs. 
Consequences of the photon-ALP interaction on the polarization of photons produced by gamma-ray bursts have been analyzed in~\cite{bassan}, and photon-ALP conversion effects on the polarization of photons originated from other astrophysical sources have been studied e.g. in~\cite{ALPpol1,ALPpol2,ALPpol3,ALPpol4,ALPpol5,day}. In addition, it has been realized that the photon-ALP interaction can be used to measure the {\it emitted} photon polarization~\cite{galantiTheorems}. New attention on this topic has been recently paid because of some existing or proposed experiments that measure the polarization of cosmic photons in the X-ray band like IXPE~\cite{ixpe}, eXTP~\cite{extp}, XL-Calibur~\cite{xcalibur}, NGXP~\cite{ngxp} and XPP~\cite{xpp} and in the high-energy (HE) band such as COSI~\cite{cosi}, e-ASTROGAM~\cite{eastrogam1,eastrogam2} and AMEGO~\cite{amego}. 

In this paper, we study the propagation of the photon-ALP beam analyzing the ALP-induced modification on observed photon polarization. We calculate the final photon survival probability $P_{\gamma \to \gamma}$ and the photon degree of linear polarization $\Pi_L$ of the photon-ALP beam while crossing different magnetized environments using state-of-the-art knowledge (see following Sections). In particular, we consider the case where photons are generated and oscillate into ALPs inside the magnetic field of the jet when a blazar is present and the alternative case of photon production in the central region of a galaxy cluster. We then study the propagation of the photon-ALP beam inside the  Kolmogorov-type turbulent galaxy cluster magnetic field. We consider both the cases of cool-core (CC) and non-cool-core (nCC) galaxy clusters. Furthermore, for the propagation inside the extragalactic space two possibilities are taken into account: a low extragalactic magnetic field strength, $B_{\rm ext}<10^{-15} \, \rm G$ and the higher value $B_{\rm ext}=1 \, \rm nG$. While in the former case the photon-ALP conversion is negligible, it is efficient in the latter. At the end, we add the photon-ALP interaction inside the magnetic field of the Milky Way. Then, we analyze the final $\Pi_L$ in order to investigate possible features indicating hints for the ALP existence. We study the behavior of $P_{\gamma \to \gamma}$ and of $\Pi_L$ of the photon-ALP beam in the energy range $(1-10^{15}) \, \rm eV$ dividing it into three bands: (i) UV-X-ray band ($10^{-3}\, \rm {\rm keV}-10^2 \, \rm keV$), (ii) HE band ($10^{-1}\, \rm {\rm MeV}-10^4 \, \rm MeV$), (iii) VHE band ($10^{-2}\, \rm {\rm TeV}-10^3 \, \rm TeV$). While our findings can be tested by current and planned observatories in the X-ray and HE range, our results about the VHE band are nowadays only theoretical since present technologies are currently unable to detect photon polarization up to so high energies.

The paper is organized as follows. In Sec. II we briefly introduce ALPs and the photon-ALP system, while in Sec. III we deal with polarization and recall some results linking conversion/survival probability and particle polarization. Then, in Sec. IV we discuss the photon-ALP beam propagation crossing different magnetized media, in Sec. V we present our results in the three considered energy ranges, in Sec. VI we discuss our findings, while in Sec. VII we draw our conclusions.

\section{Axion-like particles}

ALPs are spin-zero, neutral, and extremely light pseudo-scalar bosons interacting primarily with photons (interactions with fermions are subdominant and therefore safely negligible) through the Lagrangian:
\begin{eqnarray}
&\displaystyle {\cal L}_{\rm ALP} =  \frac{1}{2} \, \partial^{\mu} a \, \partial_{\mu} a - \frac{1}{2} \, m_a^2 \, a^2 - \, \frac{1}{4 } g_{a\gamma\gamma} \, F_{\mu\nu} \tilde{F}^{\mu\nu} a \nonumber \\
&\displaystyle = \frac{1}{2} \, \partial^{\mu} a \, \partial_{\mu} a - \frac{1}{2} \, m_a^2 \, a^2 + g_{a\gamma\gamma} \, {\bf E} \cdot {\bf B}~a~,
\label{lagr}
\end{eqnarray}
where $a$ represents the ALP field, $F_{\mu\nu}$ is the electromagnetic tensor whose electric and magnetic components are ${\bf E}$ and ${\bf B}$, respectively and $\tilde{F}^{\mu\nu}$ is the $F_{\mu\nu}$ dual.
Concerning the photon-ALP coupling $g_{a \gamma \gamma}$ and the ALP mass $m_a$ many bounds exist in the literature such as those derived in~\cite{cast,straniero,fermi2016,payez2015,berg,conlonLim,meyer2020,limFabian,limJulia,limKripp,limRey2,mwd}. The firmest one reads $g_{a \gamma \gamma} < 0.66 \times 10^{- 10} \, {\rm GeV}^{- 1}$ for $m_a < 0.02 \, {\rm eV}$ at the $2 \sigma$ level from no detection of ALPs from the Sun~\cite{cast}.

In the presence of a strong external magnetic field, we must also consider the photon one-loop vacuum polarization effects accounted by the Heisenberg-Euler-Weisskopf (HEW) effective Lagrangian which reads
\begin{equation}
\label{HEW}
{\cal L}_{\rm HEW} = \frac{2 \alpha^2}{45 m_e^4} \, \left[ \left({\bf E}^2 - {\bf B}^2 \right)^2 + 7 \left({\bf E} \cdot {\bf B} \right)^2 \right]~,
\end{equation}
where $\alpha$ is the fine-structure constant and $m_e$ is the electron mass~\cite{hew1, hew2, hew3}.

We study a photon-ALP beam of energy $E$ propagating in the $y$-direction and crossing a magnetized medium whose external magnetic field entering Eq.~(\ref{lagr}) is denoted by ${\bf B}$, while ${\bf E}$ pertains to a propagating photon. Since the mass matrix of the $\gamma - a$ system is off-diagonal, the propagation eigenstates differ from the interaction eigenstates, producing $\gamma \leftrightarrow a$ oscillations in a similar way as oscillations of different flavor massive neutrinos with the only difference that in the case of the photon-ALP system an external ${\bf B}$ field is necessary in order to compensate for the spin mismatch between photons and ALPs. From the form of the photon-ALP coupling in Eq.~(\ref{lagr}), we infer that $a$ couples only with the component ${\bf B}_T$ of ${\bf B}$ which is transverse to the photon momentum $\bf k$ (see also~\cite{dgr2011}). The photon-ALP beam propagation equation following from ${\cal L}_{\rm ALP}$ of Eq.~(\ref{lagr}) reads
\begin{equation}
\label{propeq} 
\left(i \, \frac{d}{d y} + E +  {\cal M} (E,y) \right)  \psi(y)= 0~,
\end{equation}
with
\begin{equation}
\label{psi} 
\psi(y)=\left(\begin{array}{c}A_x (y) \\ A_z (y) \\ a (y) \end{array}\right)~,
\end{equation}
where ${\cal M} (E,y)$ represents the photon-ALP mixing matrix, while $A_x (y)$ and $A_z (y)$ are the two photon linear polarization amplitudes along the $x$ and $z$ axis, respectively and $a (y)$ denotes the ALP amplitude. In Eq.~(\ref{propeq}) we have employed the short-wavelength approximation~\cite{raffeltstodolsky}, which stands since we are working in the regime $E \gg m_a$ (as it will be clear in the following because of the chosen parameters). As a consequence, the photon-ALP beam propagation equation becomes a Schr\"odinger-like equation with the coordinate $y$ along the beam in place of the time $t$: thus, the relativistic beam can formally be treated as a three-level nonrelativistic quantum system.   

Denoting by ${\cal U}$ the {\it transfer matrix} of the photon-ALP beam propagation equation, which is the solution of Eq.~(\ref{propeq}) with initial condition ${\cal U}(E;y_0,y_0)=1$, a generic wave function possesses solution 
\begin{equation}
\label{psi2} 
\psi(y)={\cal U}(E;y,y_0)\psi(y_0)~,
\end{equation}
with $y_0$ accounting for the initial position of the beam. In the case of a non-polarized beam we have to use the polarization density matrix $\rho(y)$ satisfying the Von Neumann-like equation linked to Eq.~(\ref{propeq}), which reads
\begin{equation}
\label{vneum}
i \frac{d \rho (y)}{d y} = \rho (y) \, {\cal M}^{\dag} ( E, y) - {\cal M} ( E, y) \, \rho (y)~,
\end{equation}
whose solutions can be represented in terms of ${\cal U} \bigl( E; y, y_0 \bigr)$ as
\begin{equation}
\label{unptrmatr}
\rho ( y ) = {\cal U} \bigl(E; y, y_0 \bigr) \, \rho_0 \, {\cal U}^{\dag} \bigl(E; y, y_0 \bigr)~.
\end{equation}
Hence, the probability describing a beam in the initial state $\rho_0$ at position $y_0$ and in the final state $\rho$ at position $y$ reads
\begin{equation}
\label{unpprob}
P_{\rho_0 \to \rho} (E,y) = {\rm Tr} \Bigl[\rho \, {\cal U} (E; y, y_0) \, \rho_0 \, {\cal U}^{\dag} (E; y, y_0) \Bigr]~,
\end{equation}
with ${\rm Tr} \, \rho_0 = {\rm Tr} \, \rho =1$~\cite{dgr2011}.

By defining $\phi$ the angle that ${\bf B}_T$ forms with the $z$ axis, the mixing matrix $\cal M$ in Eq.~(\ref{propeq}) can be written as
\begin{eqnarray}
\label{mixmat}
&\displaystyle{\cal M} (E,y) \equiv \,\,\,\,\,\,\,\,\,\,\,\,\,\,\,\,\,\,\,\,\,\,\,\,\,\,\,\,\,\,\,\,\,\,\,\,\,\,\,\,\,\,\,\,\,\,\,\,\,\,\,\,\,\,\,\,\,\,\,\,\,\,\,\,\,\,\,\,\,\,\,\,\,\,\,\,\,\,\,\,\,\,\,\,\,\,\,\,\,\,\,\,\,\,\,\,\,\,\,\,\,\,\,\,\,\, \nonumber \\
&\displaystyle \left(
\begin{array}{ccc}
\Delta_{xx} (E,y) & \Delta_{xz} (E,y) & \Delta_{a \gamma}(y) \, {\rm sin} \, \phi \\
\Delta_{zx} (E,y) & \Delta_{zz} (E,y) & \Delta_{a \gamma}(y) \, {\rm cos} \, \phi \\
\Delta_{a \gamma}(y) \, {\rm sin}  \, \phi & \Delta_{ a \gamma}(y) \, {\rm cos} \, \phi & \Delta_{a a} (E) \\
\end{array}
\right)~,
\end{eqnarray}
with
\begin{equation}
\label{deltaxx}
\Delta_{xx} (E,y) \equiv \Delta_{\bot} (E,y) \, {\rm cos}^2 \, \phi + \Delta_{\parallel} (E,y) \, {\rm sin}^2 \, \phi~,
\end{equation}
\begin{eqnarray}
&\displaystyle \Delta_{xz} (E,y) = \Delta_{zx} (E,y) \equiv  \nonumber \\
&\displaystyle \left(\Delta_{\parallel} (E,y) - \Delta_{\bot} (E,y) \right) {\rm sin} \, \phi \, {\rm cos} \, \phi~,
\label{deltaxz}
\end{eqnarray}
\begin{equation}
\label{deltazz}
\Delta_{zz} (E,y) \equiv \Delta_{\bot} (E,y) \, {\rm sin}^2 \, \phi + \Delta_{\parallel} (E,y) \, {\rm cos}^2 \, \phi~,
\end{equation}
\begin{equation}
\label{deltamix} 
\Delta_{a \gamma}(y) = \frac{1}{2}g_{a\gamma\gamma}B_T(y)~,
\end{equation}
\begin{equation}
\label{deltaM} 
\Delta_{aa} (E) = - \frac{m_a^2}{2 E}~,
\end{equation}
and
\begin{eqnarray}
\label{deltaort} 
&\displaystyle \Delta_{\bot} (E,y) = \frac{i}{2 \, \lambda_{\gamma} (E,y)} - \frac{\omega^2_{\rm pl}(y)}{2 E}  \nonumber \\
&\displaystyle + \frac{2 \alpha}{45 \pi} \left(\frac{B_T(y)}{B_{{\rm cr}}} \right)^2 E + \rho_{\rm CMB}E~,
\end{eqnarray}
\begin{eqnarray}
\label{deltapar} 
&\displaystyle \Delta_{\parallel} (E,y) = \frac{i}{2 \, \lambda_{\gamma} (E,y)} - \frac{\omega^2_{\rm pl}(y)}{2 E} \nonumber \\
&\displaystyle + \frac{7 \alpha}{90 \pi} \left(\frac{B_T(y)}{B_{{\rm cr}}} \right)^2 E + \rho_{\rm CMB}E ~,    
\end{eqnarray}
where $B_{{\rm cr}} \simeq 4.41 \times 10^{13} \, {\rm G}$ is the critical magnetic field and $\rho_{\rm CMB}=0.522 \times 10^{-42}$. Eq.~(\ref{deltamix}) accounts for the photon-ALP interaction, while Eq.~(\ref{deltaM}) describes the ALP mass effect. The first term in Eqs.~(\ref{deltaort}) and~(\ref{deltapar}) accounts for absorption (e.g. due to the EBL) and $\lambda_{\gamma}$ is the $\gamma\gamma \to e^+e^-$ mean free path~\cite{foot1}. In the second term of Eqs.~(\ref{deltaort}) and~(\ref{deltapar}) $\omega_{\rm pl}$ is the plasma frequency, which is related to the electron number density $n_e$ by $\omega_{\rm pl}=(4 \pi \alpha n_e / m_e)^{1/2}$. The third term in Eqs.~(\ref{deltaort}) and~(\ref{deltapar}) accounts for the photon one-loop vacuum polarization coming from ${\cal L}_{\rm HEW}$ of Eq.~(\ref{HEW}) and which produces polarization variation and birifrangence on the beam, while the fourth term represents the contribution from photon dispersion on the cosmic microwave background (CMB)~\cite{raffelt2015} which produces sizable effects inside the extragalactic space~\cite{grExt}.

In order to understand the different regimes defined by the relative importance of the $\Delta$ terms in Eq.~(\ref{mixmat}) that the photon-ALP system can experience, we consider the case of: (i) fully polarized photons, (ii) no photon absorption i.e. $\lambda_{\gamma} \to \infty$, (iii) homogeneous medium, (iv) constant $\bf B$ field, so that ${\bf B}(y)\equiv{\bf B}, \forall \, y$ having thus the freedom to choose the $z$ axis along the direction of ${\bf B}_T$ -- this fact translates to set $\phi=0$ in Eq.~(\ref{mixmat}). With these assumptions the $\gamma \to a$ conversion probability reads
\begin{equation}
\label{convprob}
P_{\gamma \to a} (E, y) = \left(\frac{g_{a\gamma\gamma}B_T \, l_{\rm osc} (E)}{2\pi} \right)^2 {\rm sin}^2 \left(\frac{\pi (y-y_0)}{l_{\rm osc} (E)} \right)~,
\end{equation}
where
\begin{equation}
\label{losc}     
l_{\rm osc} (E) \equiv \frac{2 \pi}{\left[\bigl(\Delta_{zz} (E) - \Delta_{aa} (E) \bigr)^2 + 4 \, \Delta_{a\gamma}^2 \right]^{1/2}}~
\end{equation}
is the photon-ALP beam oscillation length. It is useful to define the {\it low-energy threshold} $E_L$ and the {\it high-energy threshold} $E_H$ as
\begin{equation}
\label{EL}
E_L \equiv \frac{|m_a^2 - \omega^2_{\rm pl}|}{2 g_{a \gamma \gamma} \, B_T}~,  
\end{equation}
and 
\begin{equation}
\label{EH}
E_H \equiv g_{a \gamma \gamma} \, B_T \left[\frac{7 \alpha}{90 \pi} \left(\frac{B_T}{B_{\rm cr}} \right)^2 + \rho_{\rm CMB} \right]^{- 1}~,
\end{equation} 
respectively. For $E_L \lesssim E \lesssim E_H$ the {\it strong-mixing} regime takes place and the plasma contribution, the ALP mass term, the QED one-loop effect and the photon dispersion on the CMB are negligible. In such a situation the $P_{\gamma \to a}$ is maximal, energy independent and reads
\begin{equation}
\label{convprobSM}
P_{\gamma \to a} (y) = {\rm sin}^2 \left( \frac{g_{a\gamma\gamma} B_T}{2}  (y-y_0) \right)~.
\end{equation}
For $E \lesssim E_L$ the plasma contribution and/or the ALP mass term dominate and the same is true for $E \gtrsim E_H$ concerning the QED one-loop effect and/or the photon dispersion on the CMB: in both the cases we are in the {\it weak-mixing} regime and $P_{\gamma \to a}$ becomes energy dependent and progressively vanishes.

Everything we have discussed above in the case of fully polarized photons, no absorption and homogeneous and constant $\bf B$ field can be translated in the general case: however, the analytic expressions of the equations would be unacceptably cumbersome and would shed no light on what is going on, so that we have decided to report the considered simplified case. In the following Sections the appropriate photon polarization, photon absorption (if present) and the complete spatial-dependent expressions of $\bf B$ and $n_e$ are considered.

\section{Polarization effects}
Whenever the polarization of the photon-ALP beam is not measurable -- as in the VHE band -- or the beam is expected to be unpolarized, the generalized {\it polarization density matrix} $\rho$ must be used: the matrix $\rho$ associated to the photon-ALP beam can be written as 
\begin{equation}
\label{densmat}
\rho (y) = \left(\begin{array}{c}A_x (y) \\ A_z (y) \\ a (y)
\end{array}\right)
\otimes \left(\begin{array}{c}A_x (y) \  A_z (y) \ a (y) \end{array}\right)^{*}~,
\end{equation}
which allows to treat unpolarized, partially-polarized and totally polarized beams (pure states), at once. Pure photon states in the $x$ and $z$ direction read
\begin{equation}
\label{densphot}
{\rho}_x = \left(
\begin{array}{ccc}
1 & 0 & 0 \\
0 & 0 & 0 \\
0 & 0 & 0 \\
\end{array}
\right)~, \,\,\,\,\,\,\,\,
{\rho}_z = \left(
\begin{array}{ccc}
0 & 0 & 0 \\
0 & 1 & 0 \\
0 & 0 & 0 \\
\end{array}
\right)~,
\end{equation}
the ALP state can be expressed by
\begin{equation}
\label{densa}
{\rho}_a = \left(
\begin{array}{ccc}
0 & 0 & 0 \\
0 & 0 & 0 \\
0 & 0 & 1 \\
\end{array}
\right)~,
\end{equation}
while unpolarized photons are described by
\begin{equation}
\label{densunpol}
{\rho}_{\rm unpol} = \frac{1}{2} \left(
\begin{array}{ccc}
1 & 0 & 0 \\
0 & 1 & 0 \\
0 & 0 & 0 \\
\end{array}
\right)~.
\end{equation}
Partially polarized photons are associated to a polarization density matrix that possesses an intermediate functional expression between Eqs.~(\ref{densphot}) and Eq.~(\ref{densunpol}). 

We can express the $2 \times 2$ photon polarization density matrix -- which is the 1-2 submatrix of the polarization density matrix of the photon-ALP system of Eq.~(\ref{densmat}) -- in terms of the Stokes parameters as~\cite{poltheor1}
\begin{equation}
\label{stokes}
{\rho}_{\gamma} = \frac{1}{2} \left(
\begin{array}{cc}
I+Q & U-iV \\
U+iV & I-Q \\
\end{array}
\right)~,
\end{equation}
while the definition of the photon degree of {\it linear polarization} $\Pi_L$ reads~\cite{poltheor2}
\begin{equation}
\label{PiL}
\Pi_L \equiv \frac{(Q^2+U^2)^{1/2}}{I}~,
\end{equation}
which in terms of the photon polarization density matrix elements $\rho_{ij}$ with $i,j=1,2$ can be expressed as
\begin{equation}
\label{PiL}
\Pi_L = \frac{\left[ (\rho_{11}-\rho_{22})^2+(\rho_{12}+\rho_{21})^2\right]^{1/2}}{\rho_{11}+\rho_{22}}~.
\end{equation}
The photon-ALP interaction induces a sizable modification on the final $\Pi_L$ with respect to the initial photon degree of linear polarization $\Pi_{L,0}$, as it will be evident in the following Sections.

We want to conclude this Section by recalling some results linking conversion/survival probability and {\it initial} particle polarization which come from some theorems enunciated and demonstrated in~\cite{galantiTheorems}. We verify that our results about the photon survival probability $P_{\gamma \to \gamma}$ and reported in the figures of Sec. V satisfy such theorems.

\begin{enumerate}[(i)]

\item In any isolated system consisting of photons interacting with ALPs only, where photons are not absorbed and with initial condition of only photons with {\it initial} degree of linear polarization $\Pi_{L,0}$, the conversion probability satisfies the inequality $P_{\gamma \to a} \le (1+\Pi_{L,0})/2$, while $P_{\gamma \to \gamma} \ge (1-\Pi_{L,0})/2$.

\item In the previous conditions but in the case of initially unpolarized photons ($\Pi_{L,0}=0$), we observe $P_{\gamma \to a} \le 1/2$ and $P_{\gamma \to \gamma} \ge 1/2$.

\item In the conditions of item (i) $\Pi_{L,0}$ represents the measure of the overlap between the values assumed by $P_{\gamma \to a}$ and $P_{\gamma \to \gamma}$.

\item In the conditions of item (ii) $\Pi_{L,0}=0$ establishes that $P_{\gamma \to a}$ and $P_{\gamma \to \gamma}$ possess the common value of 1/2, at most.

\end{enumerate}

\section{Photon-ALP beam propagation}

In this Section we analyze the photon-ALP beam propagation in all the media considered in this paper: the blazar jet, the host galaxy, the galaxy cluster, the extragalactic space and the Milky Way. Hereafter, we cursorily recall the main properties and consequences of the photon-ALP beam propagation in such media letting the details to the specific papers cited below, which are dedicated to that particular subject. About the photon-ALP interaction parameters we take $g_{a\gamma\gamma}=0.5 \times 10^{-11} \, \rm GeV^{-1}$ and two cases concerning the ALP mass: (i) $m_a \lesssim 10^{-14} \, \rm eV$, (ii) $m_a = 10^{-10} \, \rm eV$ (more about this, later). These choices allow us to stay within the current firmest bound (see Sec. V for more details). In any case, we want to stress that all existing bounds about $g_{a\gamma\gamma}$ and $m_a$ have to be viewed as indications at most and even a choice of such parameters beyond these limits is perfectly allowed.

\subsection{Active galactic nuclei}

Active galactic nuclei (AGN) are basically extragalactic supermassive black holes (SMBHs) accreting matter from neighborhood and in which two collimated relativistic jets develop in opposite directions. When one of the jets occasionally points towards us, AGN are called blazars. Blazars are divided into two groups: flat spectrum radio quasars (FSRQs) and BL Lac objects (BL Lacs). While FSRQs are more powerful and characterized by strong optical emission lines and by the presence of high absorption zones for VHE photons (broad line region, torus; see e.g.~\cite{torusCTA}), BL Lacs are less powerful and possess neither sizable emission lines nor the above-mentioned absorption regions. BL Lacs have a harder spectrum reaching observed energies up to $\sim 20 \, \rm TeV$ for close sources (see e.g. Markarian 501~\cite{hegra}). We consider BL Lacs in this paper.

By closely following the results obtained in~\cite{trg2015}, we study here the propagation of the photon-ALP beam inside the magnetic field ${\bf B}^{\rm jet}$ of the jet. 
We start from the photon emission region placed at a distance of about $y_{\rm em}= (10^{16} - 10^{17}) \, {\rm cm}$ from the central SMBH -- for definiteness we take $y_{\rm em}= 3 \times 10^{16} \, {\rm cm}$ -- up to the distance where the jet ends at about $1 \, \rm kpc$, entering the host galaxy. Concerning ${\bf B}^{\rm jet}$ what is relevant is its toroidal part which is transverse to the jet axis~\cite{bbr1984,ghisellini2009,pudritz2011}. Its profile reads
\begin{equation}
\label{Bjet}
B^{\rm jet} ( y ) = B^{\rm jet}_0 \left(\frac{y_{{\rm em}}}{y}\right)~,
\end{equation}
where $B^{\rm jet}_0$ is the jet magnetic field strength at the photon emission position $y_{\rm em}$. Because of the conical shape of the jet, the electron number density $n_e^{\rm jet}$ profile is expected to be represented by
\begin{equation}
\label{njet}
n^{\rm jet}_e ( y ) = n^{\rm jet}_{e,0} \left(\frac{y_{{\rm em}}}{y}\right)^2~,
\end{equation}
where $n^{\rm jet}_{e,0}$ is the jet electron number density at $y_{\rm em}$. Synchrotron Self Compton (SSC) diagnostics applied to blazar spectra can give information about realistic values for $B^{\rm jet}_0$ and $n^{\rm jet}_{e,0}$~\cite{tavecchio2010}. For definiteness, we take the average values $B^{\rm jet}_0=0.5 \, \rm G$ and $n^{\rm jet}_{e,0}=5 \times 10^4 \, \rm cm^{-3}$.

Once all the above quantities are fixed, the whole propagation process of the photon-ALP beam within the jet can be evaluated and we can calculate its transfer matrix ${\cal U}_{\rm jet}$ (for more details see~\cite{trg2015}). 

By denoting the Lorentz factor with $\gamma$, since we calculate the photon-ALP beam propagation crossing the jet in its comoving frame, we must apply the transformation $E \to \gamma E$ to the beam in order to translate it to the fixed frames of the following regions. We take $\gamma=15$.

\subsection{Host galaxy}

BL Lacs are normally located in elliptical galaxies, where the magnetic field ${\bf B}_{\rm host}$ is believed to be of turbulent nature. A domain-like model is commonly used to describe the ${\bf B}_{\rm host}$ behavior, while its average strength and coherence length are $B_{\rm host} \simeq 5 \, \mu{\rm G}$ and $L_{\rm dom}^{\rm host} \simeq 150 \, {\rm pc}$, respectively~\cite{moss1996}.

Since the $\gamma \leftrightarrow a$ oscillation length is much larger than $L_{\rm dom}^{\rm host}$, the photon-ALP conversion turns out to be totally inefficient in this region, so that the effect of the host galaxy on the whole photon-ALP beam propagation process is subdominant, as shown in~\cite{trgb2012}. 
Yet, we carefully calculate the transfer matrix in the host galaxy ${\cal U}_{\rm host}$. 

\subsection{Galaxy cluster}

Faraday rotation measurements and synchrotron radio emissions establish the existence of ${\cal O}(1-10) \, \mu{\rm G}$ magnetic fields ${\bf B}^{\rm clu}$ inside galaxy clusters~\cite{cluB1,cluB2}. While old models described ${\bf B}^{\rm clu}$ with a domain-like structure, a better characterization of ${\bf B}^{\rm clu}$ is nowadays established. In particular, ${\bf B}^{\rm clu}$ is of isotropic gaussian turbulent nature and possesses a Kolmogorov-type turbulence power spectrum $M(k)\propto k^q$ with $k$ the wave number in the interval $[k_L,k_H]$ and index $q=-11/3$~\cite{cluFeretti}. For definiteness, we take $k_L=0.2 \, \rm kpc^{-1}$ and $k_H=3 \, \rm kpc^{-1}$. The behavior of ${\bf B}^{\rm clu}$ and of the cluster electron number density $n_e^{\rm clu}$ with respect to the radial distance reads~\cite{betaModel,cluFeretti,clu2}
\begin{equation}
\label{eq1}
B^{\rm clu}(y)={\cal B} \left( B_0^{\rm clu},k,q,y \right) \left( \frac{n_e^{\rm clu}(y)}{n_{e,0}^{\rm clu}} \right)^{\eta_{\rm clu}}~,
\end{equation}
and
\begin{equation}
\label{eq2}
n_e^{\rm clu}(y)=n_{e,0}^{\rm clu} \left( 1+\frac{y^2}{r_{\rm core}^2} \right)^{-\frac{3}{2}\beta_{\rm clu}}~,
\end{equation}
respectively, where ${\cal B}$ represents the spectral function accounting for the Kolmogorov-type turbulence of the cluster magnetic field (see e.g.~\cite{meyerKolm} for more details), $B_0^{\rm clu}$ and $n_{e,0}^{\rm clu}$ are the central cluster magnetic field strength and the central electron number density, respectively, while $\eta_{\rm clu}$ and $\beta_{\rm clu}$ are two parameters of the cluster and $r_{\rm core}$ is the cluster core radius. In the following, we employ average values for the above cluster parameters, by considering $B_0^{\rm clu}=15 \, \mu{\rm G}$, $\eta_{\rm clu}=0.75$ and the typical values $\beta_{\rm clu}=2/3$ and $r_{\rm core}=100 \, \rm kpc$~\cite{cluFeretti,clu2,cluValues}.

The choice of the value of $n_{e,0}^{\rm clu}$ is more involved. Two main categories of galaxy clusters exist: cool-core (CC) and non-cool-core (nCC) galaxy clusters (see also note~\cite{noteClu}). While CC galaxy clusters usually host an AGN, the SMBH in the center of nCC galaxy clusters is generally not active. Some studies propose an interplay between active/quiescent SMBHs and CC/nCC galaxy clusters suggesting that the two systems are linked and with the one influencing the evolution of the other~\cite{cluEvol}. Although CC and nCC galaxy clusters differ in many aspects (see e.g.~\cite{cluValues}), what is important for our studies is their central electron number density $n_{e,0}^{\rm clu}$. We consider $n_{e,0}^{\rm clu}= 5 \times 10^{-2} \, \rm cm^{-3}$ for CC galaxy clusters and $n_{e,0}^{\rm clu}=0.5 \times 10^{-2} \, \rm cm^{-3}$ for nCC ones, which represent the average values for the two classes~\cite{cluValues}.

\begin{figure}
\centering
\includegraphics[width=0.45\textwidth]{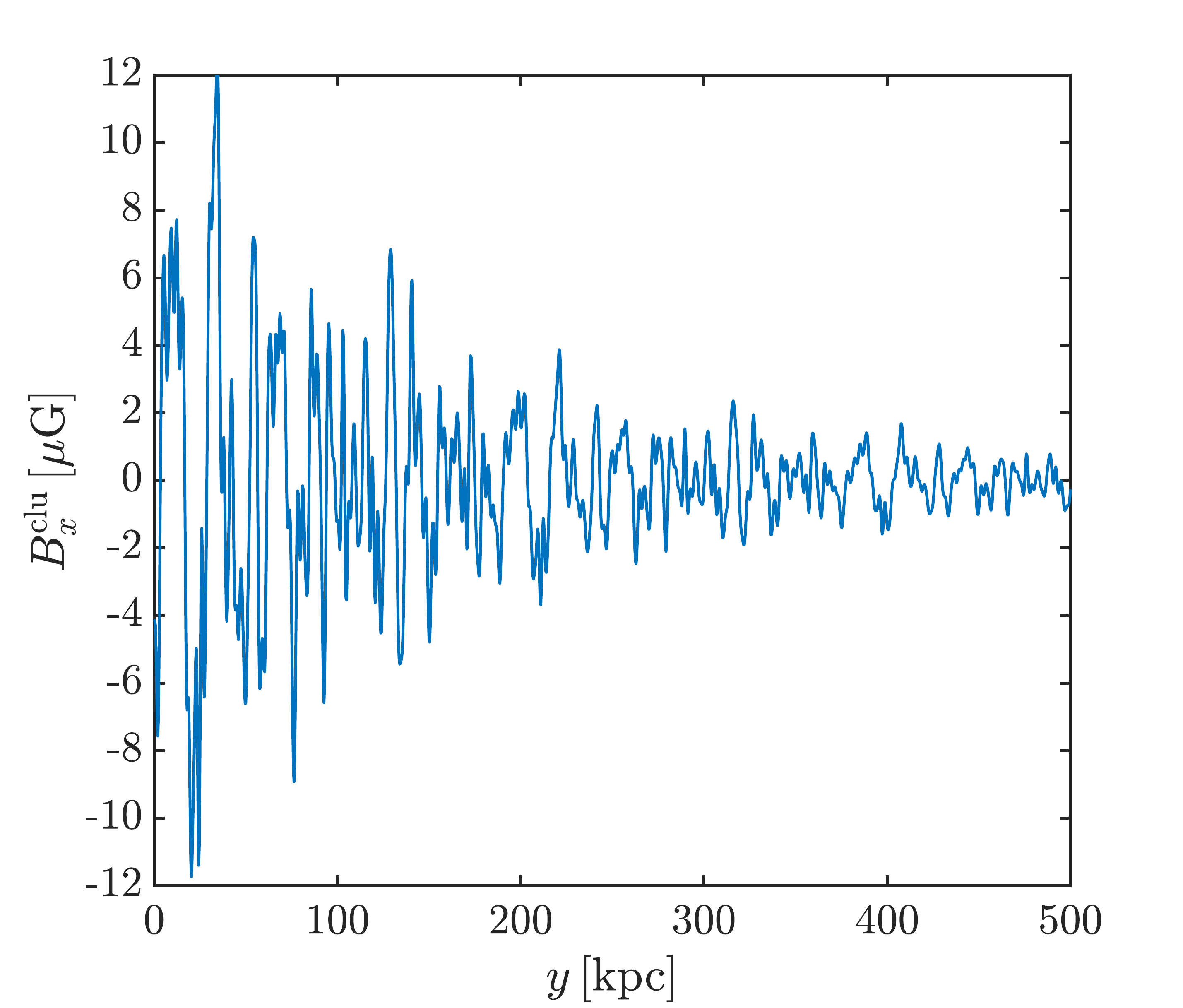}
\caption{\label{Bclu} Component along the $x$-axis of the galaxy cluster turbulent magnetic field $B^{\rm clu}_x$ with respect to the cluster radial distance $y$.}
\end{figure}

As an example, we plot a realization of the component along the $x$-axis of the galaxy cluster turbulent magnetic field $B^{\rm clu}_x$ with respect to the cluster radial distance $y$ in Fig.~\ref{Bclu} with the above-reported choice of the cluster parameters.

Then, by propagating the photon-ALP beam in the cluster starting from the central region up to its external border (we take a cluster radius of $1 \, \rm Mpc$), we obtain the transfer matrix ${\cal U}_{\rm clu}$ of the photon-ALP system inside the cluster.

\subsection{Extragalactic space}

The extragalactic magnetic field ${\bf B}_{\rm ext}$ affects the photon-ALP beam propagation in an amount which depends on the strength and morphology of ${\bf B}_{\rm ext}$. However, our knowledge of ${\bf B}_{\rm ext}$ is nowadays very poor: $B_{\rm ext}$ is restricted by current limits to the range $10^{- 7} \, {\rm nG} \leq {B}_{\rm ext} \leq 1.7 \, {\rm nG}$ on the scale of ${\cal O} (1) \, {\rm Mpc}$~\cite{neronov2010,durrerneronov,pshirkov2016}. Although several models for ${\bf B}_{\rm ext}$ exist in the literature~\cite{kronberg1994,grassorubinstein,wanglai,jap}, ${\bf B}_{\rm ext}$ is believed to possess a {\it domain-like} structure: ${\bf B}_{\rm ext}$ keeps a constant strength in each domain and the same direction over an entire domain of size $L_{\rm dom}^{\rm ext}$ which is equal to the magnetic field coherence length, but it randomly and {\it discontinuously} varies its direction crossing from one domain to the following one~\cite{kronberg1994,grassorubinstein}. By turbulence amplified outflows from primeval galaxies~\cite{reessetti,hoyle,kronberg1999,furlanettoloeb} predict values for the extragalactic magnetic field in the upper range of the existing limits: $B_{\rm ext} = {\cal O}(1) \, \rm nG$ for a coherence length equal to the size of the magnetic domains $L_{\rm dom}^{\rm ext} = {\cal O}(1) \, \rm Mpc$.

Especially in the VHE range where the $\gamma \leftrightarrow a$ oscillation length $l_{\rm osc}$ can become smaller than $L_{\rm dom}^{\rm ext}$ -- in this case we can have $E \gtrsim E_H$ because of the photon dispersion on the CMB~\cite{raffelt2015} (see Eq.~(\ref{EH}) in Sec. II) -- the simple discontinuous domain-like model for ${\bf B}_{\rm ext}$ produces {\it unphysical} results about the photon-ALP beam propagation since the system becomes sensitive to the ${\bf B}_{\rm ext}$ substructure. This is the reason why an improved {\it physically consistent} continuous domain-like model has been developed in~\cite{grSM}, where ${\bf B}_{\rm ext}$ maintains the same strength in all domains and its orientation is constant in the central part of the domain but {\it continuously and smoothly} changes direction passing -- still randomly -- from a domain to the following one. This procedure preserves the domain-like structure of ${\bf B}_{\rm ext}$ correcting the unphysical behavior at the domain edge crossing and still permits an analytical even if cumbersome solution of Eq.~(\ref{propeq})~\cite{grSM}.

Since a ${\bf B}_{\rm ext}$ high strength scenario is favored but not certain, we consider two cases in this paper: (i) $B_{\rm ext}=1 \, \rm nG$ with $L_{\rm dom}^{\rm ext}$ randomly varying according to a power-law distribution function $\propto (L_{\rm dom}^{\rm ext})^{-1.2}$ in the range $(0.2-10) \, \rm Mpc$ and with $\langle L_{\rm dom}^{\rm ext} \rangle = 2 \, \rm Mpc$ -- which is consistent with present bounds~\cite{durrerneronov}, (ii) $B_{\rm ext}<10^{-15} \, \rm G$. In the former case the photon-ALP conversion is efficient at VHE and produces sizable effects on the photon-ALP beam propagation reducing the VHE photon absorption caused by the interaction with the EBL photons~\cite{grExt}: we consider the EBL model of Franceschini and Rodighiero~\cite{franceschinirodighiero}. Instead, the photon-ALP interaction is totally negligible in the latter case, so that propagation in the extragalactic space is dominated by EBL absorption, when present. In both the previous cases we can calculate the transfer matrix of the photon-ALP beam in the extragalactic space ${\cal U}_{\rm ext}$ by following the above-discussed strategy and developed in~\cite{grSM,grExt}.

\subsection{Milky Way}

The knowledge of the Milky Way magnetic field ${\bf B}_{\rm MW}$ has greatly improved in the last years: it is well known that it possesses a strength of the order ${\cal O}(1) \, \mu{\rm G}$ and presents both a turbulent and a regular component. The regular part of ${\bf B}_{\rm MW}$ produces the dominant effects on the photon-ALP beam propagation, while the contribution of the turbulent part can often be discarded since the coherence length of the turbulent field is much smaller than the $\gamma \leftrightarrow a$ oscillation length. Nevertheless, accurate maps concerning the profile of ${\bf B}_{\rm MW}$ and its behavior with respect to the observational direction and distance nowadays exist in the literature~\cite{jansonfarrar1,jansonfarrar2,pshirkovMF2011}. For this reason in this paper we calculate the photon-ALP beam propagation inside the Milky Way by closely following the strategy developed in~\cite{gtre2019} by using the model of Jansson and Farrar~\cite{jansonfarrar1,jansonfarrar2}, which takes into account a disk and a halo component, both parallel to the Galactic plane, and a poloidal `X-shaped' component at the galactic center. In addition, newer data about polarized synchrotron and different models of the cosmic ray and thermal electron distribution are described in the newer version~\cite{uf2017}. We improve the description of the turbulent component of ${\bf B}_{\rm MW}$ by using the model developed in~\cite{BMWturb}.

We have tested that our results are qualitatively unchanged by using the model of Pshirkov {\it et al.}~\cite{pshirkovMF2011} but we have preferred the model of Jansson and Farrar~\cite{jansonfarrar1,jansonfarrar2} for our calculation since the one of Pshirkov {\it et al.}~\cite{pshirkovMF2011} does not determine the Galactic halo component of ${\bf B}_{\rm MW}$ with accuracy. The electron number density inside the Milky Way disk is $n_e^{\rm MW} \simeq 1.1 \times 10^{-2} \, {\rm cm}^{-3}$, as inferred from the model developed in~\cite{ymw2017}, which we employ in this paper.

By using the strategy developed in~\cite{gtre2019} and the model~\cite{jansonfarrar1,jansonfarrar2,BMWturb} concerning ${\bf B}_{\rm MW}$, we calculate the transfer matrix ${\cal U}_{\rm MW}$ of the photon-ALP system inside the Milky Way for a specific direction. In order to be conservative, we consider our source as placed in the direction of the Galactic pole, where ${\bf B}_{\rm MW}$ is smaller and the photon-ALP conversion is less efficient than in other directions.

\subsection{Overall photon-ALP beam propagation}

By knowing all transfer matrices in each region, we can calculate the total transfer matrix $\cal U$ of the photon-ALP system both in the case where photons are generated in the central region of a galaxy cluster with $\cal U$ reading
\begin{equation} 
\label{Utot2}
{\cal U}={\cal U}_{\rm MW}\,{\cal U}_{\rm ext}\,{\cal U}_{\rm clu}~,
\end{equation}
and in the alternative scenario of photons produced in the jet of a blazar, in which case $\cal U$ is
\begin{equation} 
\label{Utot1}
{\cal U}={\cal U}_{\rm MW}\,{\cal U}_{\rm ext}\,{\cal U}_{\rm clu}\,{\cal U}_{\rm host}\,{\cal U}_{\rm jet}~.
\end{equation}
As discussed in Sec. II, the whole final photon survival probability can be expressed as
\begin{equation} 
\label{probSurvFinal}
P_{\gamma \to \gamma}  = \sum_{i = x,z} {\rm Tr} \left[\rho_i \, {\cal U} \, \rho_{\rm in} \, {\cal U}^{\dagger} \right]~,
\end{equation}
where $\rho_x$ and $\rho_z$ read from Eqs.~(\ref{densphot}) while $\rho_{\rm in}$ is the beam initial polarization density matrix which can be in general unpolarized, partially polarized or fully polarized (see also Secs. II and III for definition and Sec. V for the chosen values). Instead, the whole final photon degree of linear polarization $\Pi_L$ is expressed by Eq.~(\ref{PiL}) where $\rho_{ij}$ are the elements of the final photon polarization density matrix $\rho$ that reads from Eq.~(\ref{unptrmatr}) with $\rho_0 = \rho_{\rm in}$ and by considering its 1-2 submatrix of $2 \times 2$ dimension.

\section{Photon survival probability and photon polarization}
In this Section we analyze the final photon survival probability $P_{\gamma \to \gamma}$ and the corresponding photon degree of linear polarization $\Pi_L$ resulting from the propagation of the photon-ALP beam crossing the several different magnetized media discussed in Sec. IV (blazar jet, host galaxy, galaxy cluster, extragalactic space and Milky Way). In particular, we consider two different scenarios: (i) photons are produced in the central region of a nCC galaxy cluster ($n_{e,0}^{\rm clu}= 0.5 \times 10^{-2} \, \rm cm^{-3}$), (ii) photons are generated at the blazar jet base and will propagate in a CC galaxy cluster ($n_{e,0}^{\rm clu}= 5 \times 10^{-2} \, \rm cm^{-3}$). In both cases we contemplate two possibilities: (i) high value of the extragalactic magnetic field with $B_{\rm ext}=1 \, \rm nG$ and consequently an efficient photon-ALP conversion in the extragalactic space, (ii) $B_{\rm ext}<10^{-15} \, \rm G$ resulting in a negligible photon-ALP interaction so that photons are actually subjected to EBL absorption only. 

As benchmark values concerning the photon-ALP system we take: $g_{a\gamma\gamma}=0.5 \times 10^{-11} \, \rm GeV^{-1}$ and the two values for the ALP mass: (i) $m_a \lesssim 10^{-14} \, \rm eV$, (ii) $m_a = 10^{-10} \, \rm eV$. In the former case ($m_a \lesssim 10^{-14} \, \rm eV$), we can see from Eqs.~(\ref{deltaM}--\ref{deltapar}) that the ALP mass term is smaller than the plasma term inside the AGN jet, in the cluster and in the Milky Way for the chosen parameters of the model and for the whole considered energy range. Instead, in the latter case ($m_a = 10^{-10} \, \rm eV$) the ALP mass term dominates over the plasma term in each region apart from the central zone of the blazar jet.

We calculate $P_{\gamma \to \gamma}$ and $\Pi_L$ along with its corresponding probability density function $f_{\Pi}$ for photons with observed energies $E_0$ in the range $1 \, {\rm eV}<E_0<10^{15} \, {\rm eV}$ with $E_0=E/(1+z)$ and $z$ being the redshift where photons are produced. In the following, we divide the above energy range into three intervals: (i) UV-X-ray band ($10^{-3}\, \rm {\rm keV}-10^2 \, \rm keV$), (ii) HE band ($10^{-1}\, \rm {\rm MeV}-10^4 \, \rm MeV$), (iii) VHE band ($10^{-2}\, \rm {\rm TeV}-10^3 \, \rm TeV$).

\subsection{UV-X-ray band}

In the energy range $(10^{-3}-10^2) \, \rm keV$, the photon-ALP beam propagates in the weak mixing regime and goes close to the strong mixing regime only in the upper part of the band if $m_a \lesssim 10^{-14} \, \rm eV$, as Eq.~(\ref{EL}) shows. If $m_a = 10^{-10} \, \rm eV$, the ALP mass term effect is very strong, so that the photon-ALP conversion is very inefficient to an extent that $P_{\gamma \to a} \to 0$. As a result, ALP-induced effects on $P_{\gamma \to \gamma}$ and on the final $\Pi_L$ are negligible for $m_a = 10^{-10} \, \rm eV$. This is the reason why we consider only the case $m_a \lesssim 10^{-14} \, \rm eV$ in this Subsection.

\begin{figure*}
\centering
\includegraphics[width=0.96\textwidth]{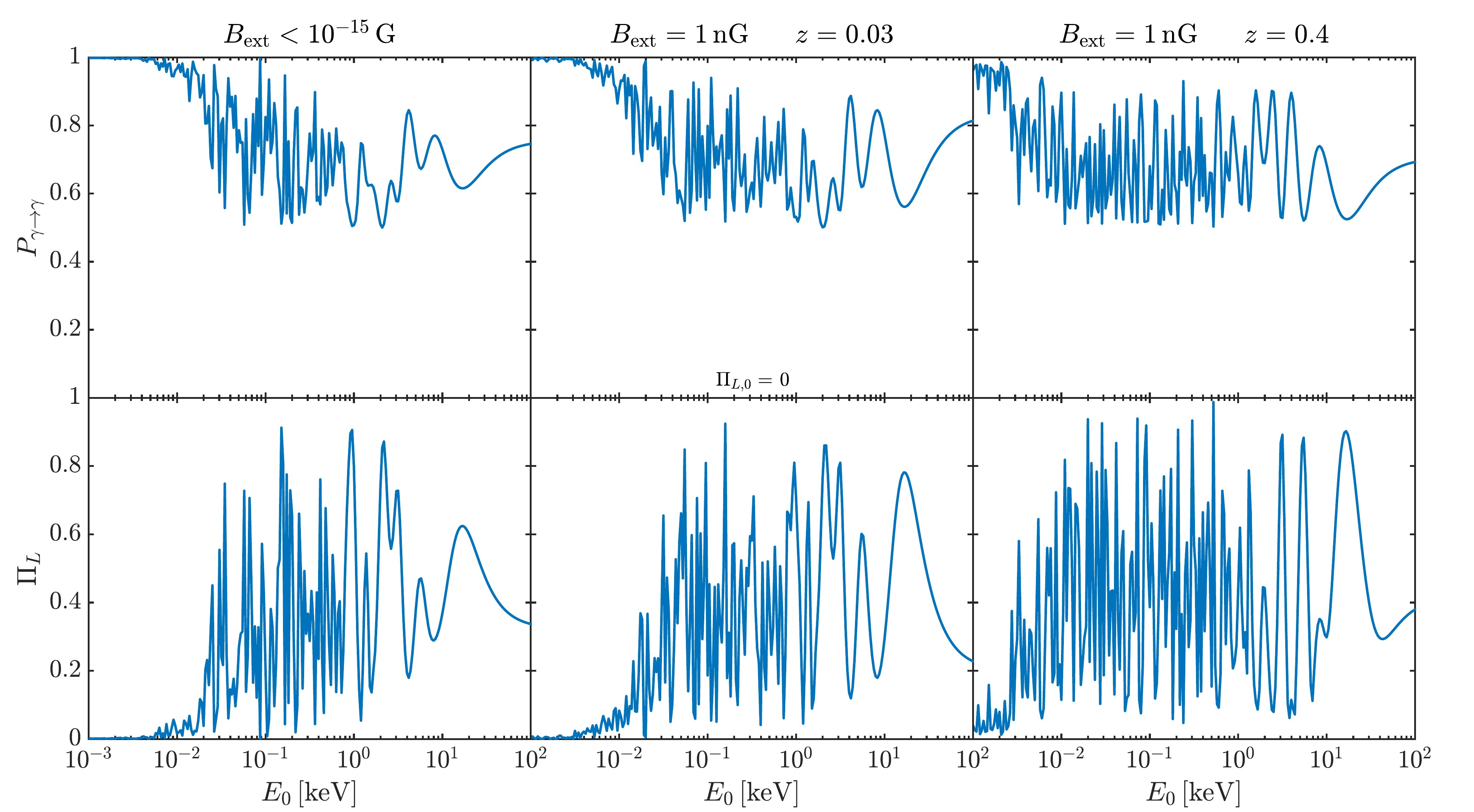}
\caption{\label{ProbPolKeV} Photon survival probability $P_{\gamma \to \gamma }$ (upper panels) and corresponding final photon degree of linear polarization $\Pi_L$ (lower panels) in the energy range $(10^{-3}-10^2) \, \rm keV$ after propagation from the cluster, where photons are produced, up to us by taking $g_{a\gamma\gamma}=0.5 \times 10^{-11} \, \rm GeV^{-1}$, $m_a \lesssim 10^{-14} \, \rm eV$ and $n_{e,0}^{\rm clu} = 0.5 \times 10^{-2} \, \rm cm^{-3}$. The initial photon degree of linear polarization is $\Pi_{L,0}=0$. In the first column an extragalactic magnetic field $B_{\rm ext} < 10^{-15} \, \rm G$ is assumed. In the second column we take $B_{\rm ext} = 1 \, \rm nG$ and a redshift $z=0.03$. In the third column we consider $B_{\rm ext} = 1 \, \rm nG$ and $z=0.4$.}
\end{figure*}

\begin{figure*}
\centering
\includegraphics[width=0.96\textwidth]{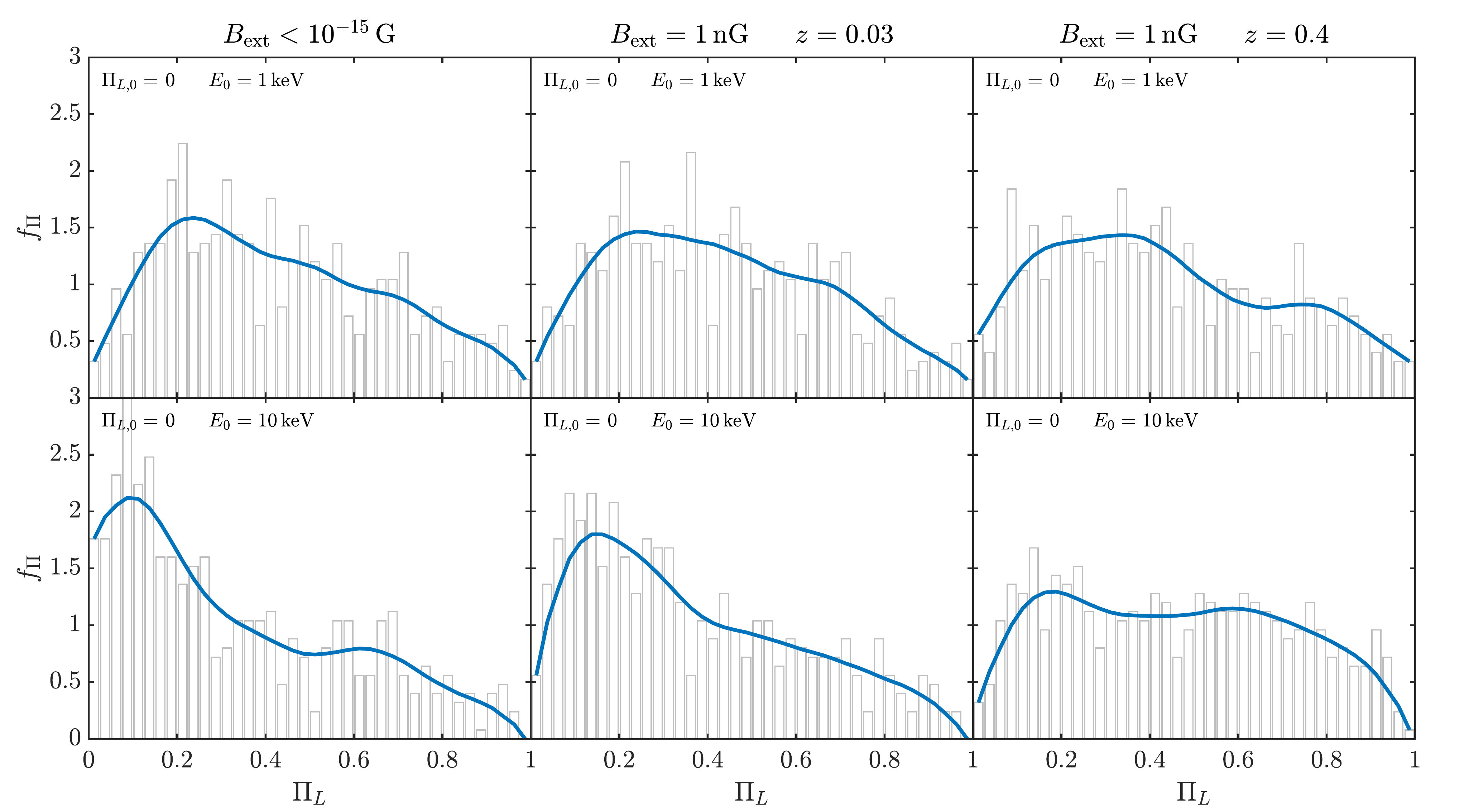}
\caption{\label{densProbKeV} Probability density function $f_{\Pi}$ arising from the plotted histogram for the final photon degree of linear polarization $\Pi_L$ at $1 \, \rm keV$ (upper panels) and $10 \, \rm keV$ (lower panels) by considering the system described in Fig.~\ref{ProbPolKeV}. The initial photon degree of linear polarization is $\Pi_{L,0}=0$. In the first column an extragalactic magnetic field $B_{\rm ext} < 10^{-15} \, \rm G$ is assumed. In the second column we take $B_{\rm ext} = 1 \, \rm nG$ and a redshift $z=0.03$. In the third column we consider $B_{\rm ext} = 1 \, \rm nG$ and $z=0.4$.}
\end{figure*}

\begin{figure*}
\centering
\includegraphics[width=0.96\textwidth]{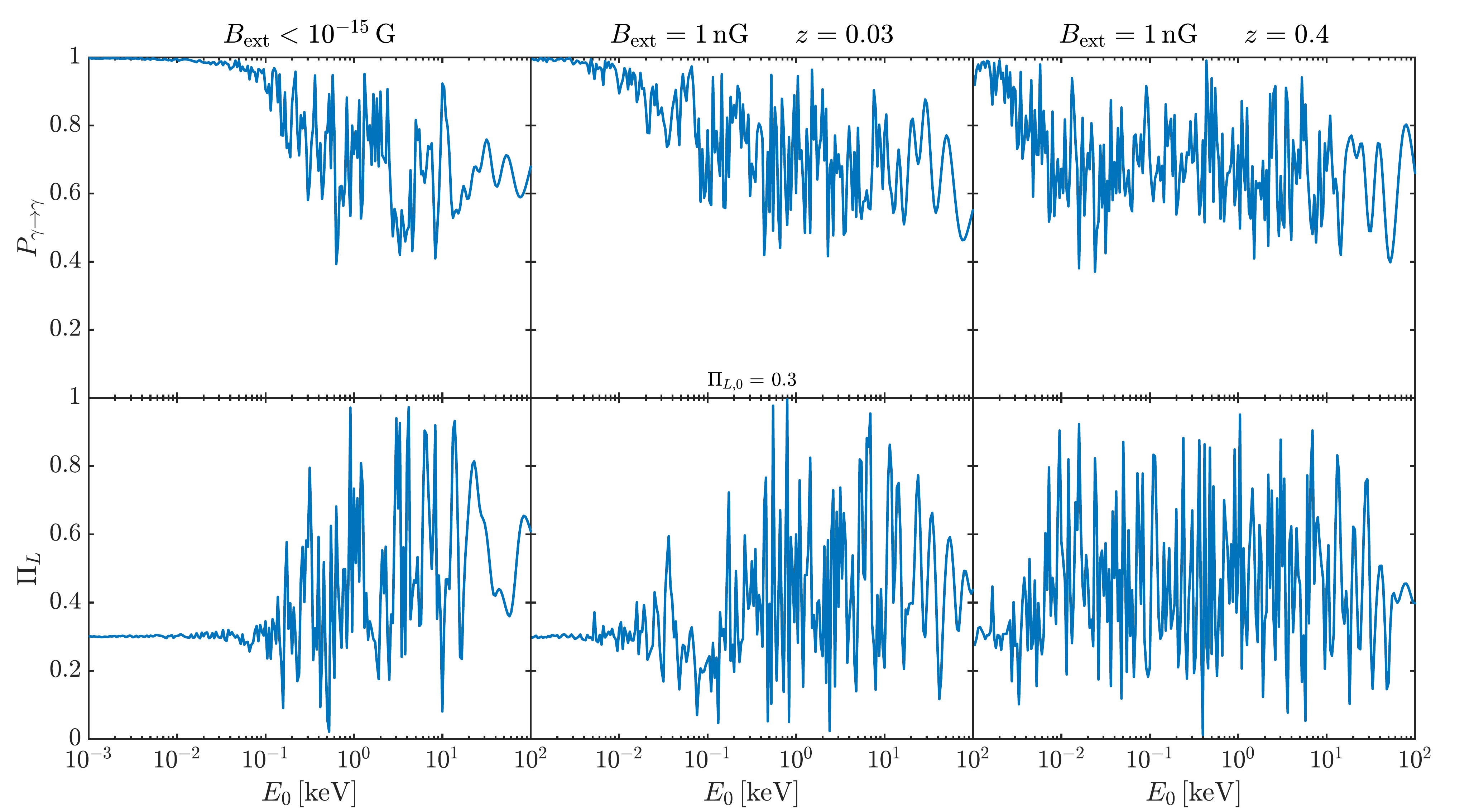}
\caption{\label{ProbPolKeVS} Same as Fig.~\ref{ProbPolKeV} but with also photon-ALP conversion within the blazar jet, where photons are produced. Thus, we accordingly take $n_{e,0}^{\rm clu} = 5 \times 10^{-2} \, \rm cm^{-3}$. The initial photon degree of linear polarization is $\Pi_{L,0}=0.3$.}
\end{figure*}

\begin{figure*}
\centering
\includegraphics[width=0.96\textwidth]{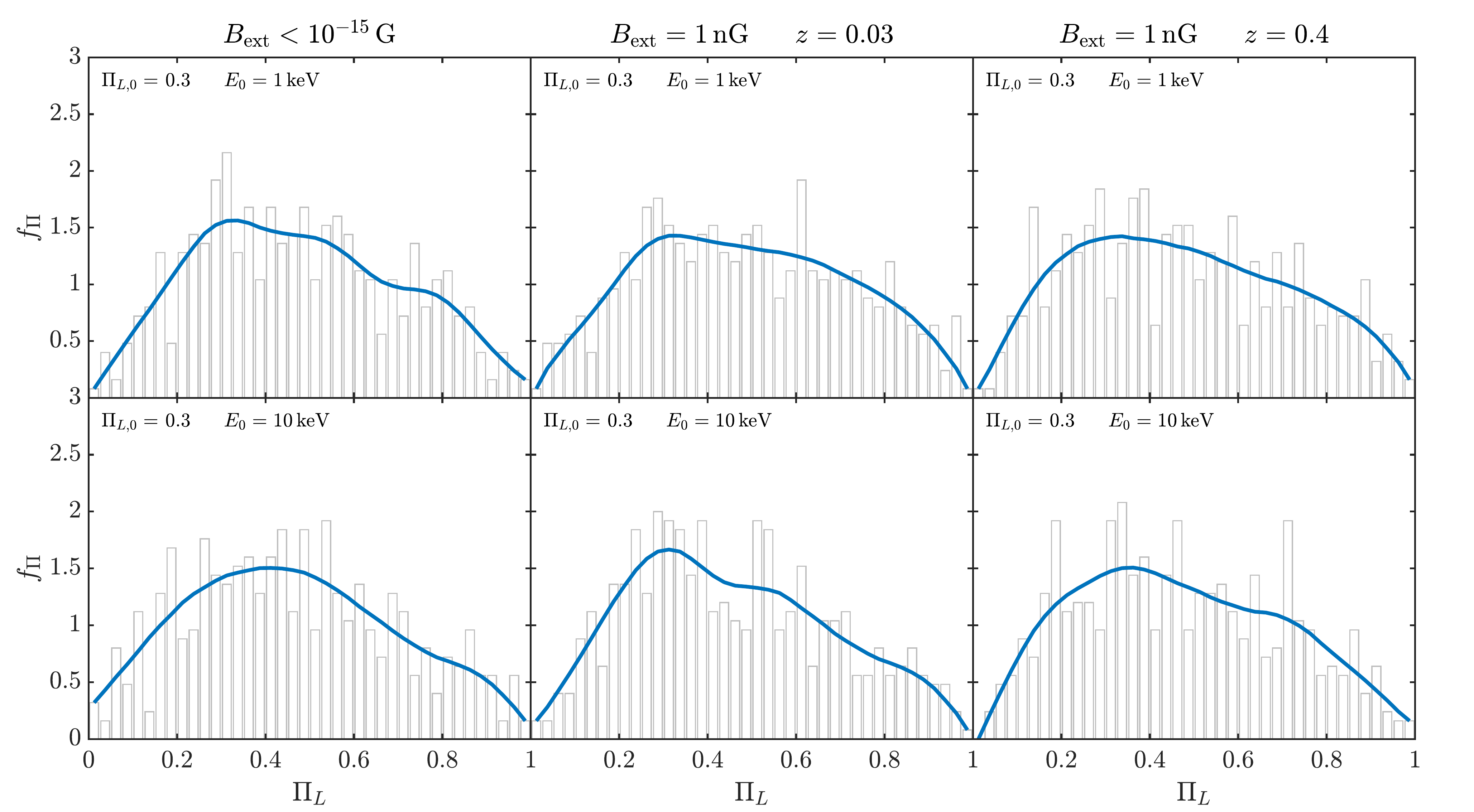}
\caption{\label{densProbKeVS} Same as Fig.~\ref{densProbKeV} but with also photon-ALP conversion within the blazar jet by considering the system described in Fig.~\ref{ProbPolKeVS}. The initial photon degree of linear polarization is $\Pi_{L,0}=0.3$.}
\end{figure*}

We show our results for the case of photons produced in the galaxy cluster in Figs.~\ref{ProbPolKeV} and~\ref{densProbKeV}, while the results concerning the alternate scenario of photons generated inside the jet of a blazar are reported in Figs.~\ref{ProbPolKeVS} and~\ref{densProbKeVS}.

We start from the case of photon production in the galaxy cluster central zone. We take $n_{e,0}^{\rm clu}= 0.5 \times 10^{-2} \, \rm cm^{-3}$ corresponding to a nCC galaxy cluster (see also Sec. IV.C). Photons are generated in the cluster central region via thermal Bremsstrahlung~\cite{mitchell1979} 
and we therefore assume them as initially unpolarized, i.e. with initial degree of linear polarization $\Pi_{L,0}=0$~\cite{poltheor2}. In Fig.~\ref{ProbPolKeV} we show $P_{\gamma \to \gamma}$ and the corresponding $\Pi_L$ in the case where the extragalactic magnetic field strength is very small with $B_{\rm ext}<10^{-15} \, \rm G$ corresponding to an inefficient photon-ALP conversion inside the extragalactic space and in the case $B_{\rm ext}=1 \, \rm nG$ with a sizable photon-ALP conversion. In the energy band considered here ($10^{-3}\, \rm {\rm keV}-10^2 \, \rm keV$), the Universe is transparent to the photon propagation with a very good accuracy, so that in the case $B_{\rm ext}<10^{-15} \, \rm G$ the transfer matrix of the photon-ALP system in the extragalactic space
reduces to ${\cal U}_{\rm ext}={\rm diag}[{\exp}(i\phi_{\gamma}),{\exp}(i\phi_{\gamma}),{\exp}(i\phi_a)]$, where $\phi_{\gamma}$ and $\phi_a$ are two phases, as there is no substantial mixing between photons and ALPs. As a result, the photon-ALP system is statistically insensible to the source redshift.

Instead, for $B_{\rm ext}=1 \, \rm nG$ the transfer matrix ${\cal U}_{\rm ext}$ remains unitary because there is still no photon absorption but 
${\cal U}_{\rm ext}$ is no more a diagonal matrix
for the increased photon-ALP conversion efficiency: now, the photon-ALP system becomes sensible to the source distance. This is the reason why we consider the two redshifts $z=0.03$ and $z=0.4$ when $B_{\rm ext}=1 \, \rm nG$.

As a general finding of Fig.~\ref{ProbPolKeV}, we observe that the weak mixing regime extends for more than three energy decades ($10^{-2}\, \rm {\rm keV}-10 \, \rm keV$): this fact reflects the big variation of the cluster magnetic field strength $B^{\rm clu}$ and electron number density $n_e^{\rm clu}$ expressed by Eq.~(\ref{eq1}) and~(\ref{eq2}), respectively, starting from their values in the cluster core up to its border. As a result, the low-energy threshold of the photon-ALP system $E_L$ of Eq.~(\ref{EL}) fails to be a single reference energy -- as would instead happen in the case of constant 
magnetic fields and electron number densities
-- and becomes an interval of energies. We can observe from the first row of Fig.~\ref{ProbPolKeV} that $P_{\gamma \to \gamma}$ never decreases down 0.5 as assured by item (ii) of Sec. III (see~\cite{galantiTheorems} for more details). In addition, from Fig.~\ref{ProbPolKeV} we infer that for $E_0 \gtrsim 10^{-2} \, \rm keV$ the photon-ALP interaction is efficient and produces sizable effects on $P_{\gamma \to \gamma}$ (see the first row of Fig.~\ref{ProbPolKeV}) and on the final $\Pi_L$ (see the second row of Fig.~\ref{ProbPolKeV}), which possess an energy-dependent behavior, as the system lies in the weak mixing regime. In addition, it seems that the case of $B_{\rm ext}<10^{-15} \, \rm G$, of $B_{\rm ext}=1 \, \rm nG$ with $z=0.03$ and of $B_{\rm ext}=1 \, \rm nG$ with $z=0.4$ are qualitatively similar.

However, this is only superficially true. In Fig.~\ref{ProbPolKeV} we plot a single realization of the photon-ALP beam propagation process, which depends on the particular realization of ${\bf B}^{\rm clu}$ and ${\bf B}_{\rm ext}$ (the variation of ${\bf B}_{\rm host}$ and ${\bf B}_{\rm MW}$ is subdominant). Since the exact orientation of these fields is unknown and only their statistical properties are, the photon-ALP beam propagation becomes a stochastic process. Nevertheless, we want to stress that what we actually observe is a single realization of the propagation process. This is the reason why we calculate several realizations of the photon-ALP beam propagation, in order to infer its statistical properties and the robustness of our results about the final $\Pi_L$. Thus, in Fig.~\ref{densProbKeV} we plot the probability density function $f_{\Pi}$ for the final $\Pi_L$ associated to the different realizations for the two benchmark energies $E_0=1 \, \rm keV$ and $E_0=10 \, \rm keV$. As a general result, we observe from Fig.~\ref{densProbKeV} that the photon-ALP interaction produces a variation of the initial $\Pi_{L,0}=0$ in all the cases and the final value $\Pi_L=0$ is never the most probable one. The effect of a high $B_{\rm ext}=1 \, \rm nG$ is to broaden $f_{\Pi}$ and to translate the expectation for the final $\Pi_L$ to larger values. The reason for this behavior lies in the stochastic nature of ${\bf B}_{\rm ext}$ and, in particular, in the distribution of the peculiar orientations of ${\bf B}_{\rm ext}$ in the several magnetic domains in a given realization (see Sec. IV.D): in the cases of close coherence when domains are crossed, the final $\Pi_L$ gets increased, while in the cases of low coherence, the final $\Pi_L$ gets decreased. This fact is more evident for $z=0.4$ with respect to $z=0.03$ since photons oscillate into ALPs longer in the former case.

In the case of photons generated inside the magnetic field of the jet of a blazar, we take $n_{e,0}^{\rm clu}= 5 \times 10^{-2} \, \rm cm^{-3}$ corresponding to a CC galaxy cluster (see also Sec. IV.C). Photons emitted at the blazar jet base in the energy range considered here ($10^{-3}\, \rm {\rm keV}-10^2 \, \rm keV$) are produced via synchrotron emission with a resulting initial polarization. Nevertheless, photons are not fully polarized and a realistic degree of linear polarization for such photons is expected to be $\Pi_{L,0}=0.2-0.4$ as discussed e.g. in~\cite{blazarPolarSincro}. Thus, we assume photons as initially partially polarized with initial degree of linear polarization $\Pi_{L,0}=0.3$. In Fig.~\ref{ProbPolKeVS} we show $P_{\gamma \to \gamma}$ and the corresponding $\Pi_L$ in the case $B_{\rm ext}<10^{-15} \, \rm G$ and when $B_{\rm ext}=1 \, \rm nG$ with the source placed at redshifts $z=0.03$ and $z=0.4$, in a similar way as we have done for the case of photon production inside the cluster. What we have previously discussed about ${\cal U}_{\rm ext}$ is still valid in this case so that we do not consider a redshift dependence in the case $B_{\rm ext}<10^{-15} \, \rm G$.

From Fig.~\ref{ProbPolKeVS} we observe that the photon-ALP beam propagates in the weak mixing regime in the interval $(10^{-2}-10) \, \rm keV$ for the same reason discussed in the case of photon production inside the cluster. Moreover, we note an additional energy dependence also in the $(10-100) \, \rm keV$ decade caused by the behavior of the blazar jet magnetic field ${\bf B}^{\rm jet}$ and of the electron number density $n_e^{\rm jet}$ of Eqs.~(\ref{Bjet}) and~(\ref{njet}), respectively: confirmation of this fact comes from Eq.~(\ref{EL}) about the value of $E_L$ in the jet. From the first row of Fig.~\ref{ProbPolKeVS} we observe that $P_{\gamma \to \gamma}$ never decreases down 0.35 as assured by item (i) of Sec. III (see~\cite{galantiTheorems} for more details). Furthermore, for $E_0 \gtrsim 10^{-2} \, \rm keV$ the first row of Fig.~\ref{ProbPolKeVS} -- where $P_{\gamma \to \gamma}$ is plotted -- shows that the photon-ALP interaction is efficient and produces sizable effects on the final $\Pi_L$ modifying the initial $\Pi_{L,0}=0.3$ (see the second row of Fig.~\ref{ProbPolKeVS}). At a first sight, the cases of $B_{\rm ext}<10^{-15} \, \rm G$, of $B_{\rm ext}=1 \, \rm nG$ with $z=0.03$ and of $B_{\rm ext}=1 \, \rm nG$ with $z=0.4$ look qualitatively similar.

What happens in the present situation is totally analogous to the case of photon production inside the cluster: thus, we calculate several realizations of the total stochastic photon-ALP beam propagation process from the blazar jet base up to the Earth and we report our results about its statistical properties in Fig.~\ref{densProbKeVS}, where we plot $f_{\Pi}$ associated to the different realizations for the two benchmark energies $E_0=1 \, \rm keV$ and $E_0=10 \, \rm keV$. Fig.~\ref{densProbKeVS} shows that the photon-ALP interaction produces a broadening of the initial $\Pi_{L,0}=0.3$ in all the cases but the final value $\Pi_L=0.3$ still remains the most probable result. For the same reasons discussed above in the case of photon generation inside the cluster, for $E_0=1 \, \rm keV$ the broadening effect on $f_{\Pi}$ increases by passing  from the case $B_{\rm ext}<10^{-15} \, \rm G$ to that $B_{\rm ext}=1 \, \rm nG$ and $z= 0.03$ and it is even more evident for $B_{\rm ext}=1 \, \rm nG$ and $z= 0.4$, while for $E_0=10 \, \rm keV$ this trend is less visible.



\subsection{High-energy band}

In the energy range $(10^{-1}-10^4) \, \rm MeV$, the $\gamma \gamma$ absorption of HE photons is totally negligible as in the UV-X-ray band, so that what we have stated above about ${\cal U}_{\rm ext}$ still holds true: thus, we have 
${\cal U}_{\rm ext}={\rm diag}[{\exp}(i\phi_{\gamma}),{\exp}(i\phi_{\gamma}),{\exp}(i\phi_a)]$
in the case $B_{\rm ext}<10^{-15} \, \rm G$, while ${\cal U}_{\rm ext}$ is still unitary but 
${\cal U}_{\rm ext}$ is no more a diagonal matrix
in the case $B_{\rm ext}=1 \, \rm nG$ for the efficiency of the photon-ALP conversion in the extragalactic space. This is the reason why only in the latter situation we consider two possibilities by placing the source (in both the cases of photon emission either in the cluster or inside the blazar jet) at redshifts $z=0.03$ and $z=0.4$. In both the cases of photons produced either inside the cluster or in the blazar jet we assume them as initially unpolarized, i.e. with initial degree of linear polarization $\Pi_{L,0}=0$. Several emission mechanisms are believed to contribute to photon production inside galaxy clusters, such as synchrotron radiation in the cluster turbulent magnetic field of electrons generated by the cascade of VHE photons, inverse Compton scattering and neutral pion decay produced in several ways~\cite{cluGammaEm1,cluGammaEm2,cluGammaEm3,cluGammaEm4} with photons emitted effectively unpolarized~\cite{polarRev,FermiPol} (see also note~\cite{footnoteClu}). Concerning the case of emission at the blazar jet base, photons are likely produced through a leptonic model via an inverse Compton process, where lower energy photons are boosted to energies in the HE and VHE bands~\cite{ic1,ic2,ic3,ic4}. Photons produced by such process are expected to be unpolarized~\cite{blazarPolarIC} (see also note~\cite{footnoteBla}).

\begin{figure*}
\centering
\includegraphics[width=0.96\textwidth]{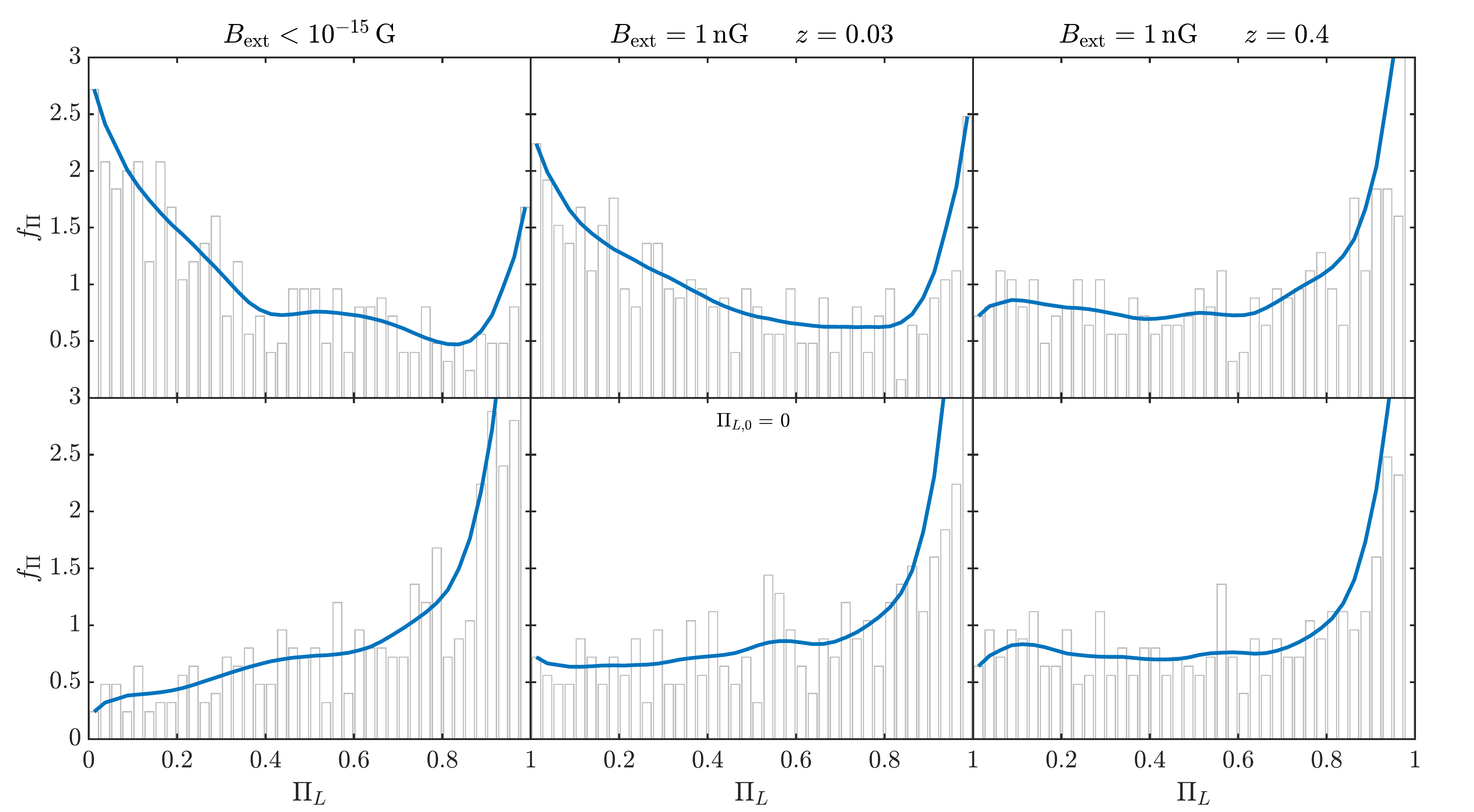}
\caption{\label{densProbHE} Probability density function $f_{\Pi}$ arising from the plotted histogram for the final photon degree of linear polarization $\Pi_L$ for photons in the energy range $(10^{-1}-10^4) \, \rm MeV$ after propagation from the emission zone up to us by taking $g_{a\gamma\gamma}=0.5 \times 10^{-11} \, \rm GeV^{-1}$ and $m_a \lesssim 10^{-14} \, \rm eV$. In the first row photons are produced in the galaxy cluster and $n_{e,0}^{\rm clu} = 0.5 \times 10^{-2} \, \rm cm^{-3}$ is assumed, while in the second row the photon-ALP beam propagates also inside the blazar jet, where photons are emitted, and we take $n_{e,0}^{\rm clu} = 5 \times 10^{-2} \, \rm cm^{-3}$. The initial photon degree of linear polarization is $\Pi_{L,0}=0$. In the first column an extragalactic magnetic field $B_{\rm ext} < 10^{-15} \, \rm G$ is assumed. In the second column we take $B_{\rm ext} = 1 \, \rm nG$ and a redshift $z=0.03$. In the third column we consider $B_{\rm ext} = 1 \, \rm nG$ and $z=0.4$.}
\end{figure*}

We start by considering the case of an ALP with mass $m_a \lesssim 10^{-14} \, \rm eV$. In such a situation, the calculation of $E_L$ from Eq.~(\ref{EL}) and of $E_H$ from Eq.~(\ref{EH}) with the parameters and corresponding profiles considered in Sec. IV concerning the magnetic field and the electron number density in the various crossed regions leads to the conclusion that the photon-ALP beam propagates in the strong mixing regime in almost the entire energy range $(10^{-1}-10^4) \, \rm MeV$ for both the cases of photon emission either in the cluster or inside the blazar jet. Since the photon-ALP system is in the strong mixing regime for $m_a \lesssim 10^{-14} \, \rm eV$, both $P_{\gamma \to \gamma}$ and $\Pi_L$ are energy independent in the HE band, so that only $f_{\Pi}$, which we plot in Fig.~\ref{densProbHE}, is really informative and gives us the statistical properties of the several realizations of the propagation process.
The strong mixing regime assures that the behavior of $f_{\Pi}$ is the same for all energies in the range $(10^{-1}-10^4) \, \rm MeV$.

From Fig.~\ref{densProbHE} we observe a general trend that is common to both the cases of photon production inside the cluster (first row, where we take $n_{e,0}^{\rm clu}= 0.5 \times 10^{-2} \, \rm cm^{-3}$ corresponding to a nCC galaxy cluster -- see also Sec. IV.C) and in the blazar jet (second row, where we assume $n_{e,0}^{\rm clu}= 5 \times 10^{-2} \, \rm cm^{-3}$ corresponding to a CC galaxy cluster -- see also Sec. IV.C) and to the various choices of $B_{\rm ext}$ and redshifts. In particular, since the system is in the strong mixing regime, the conversion probability is maximal and this fact produces a sizable modification of the final $\Pi_L$ with respect to the initial $\Pi_{L,0}=0$. For all the cases except that of photons produced in the cluster and with $B_{\rm ext}<10^{-15} \, \rm G$, the most probable values for the final $\Pi_L$ turn out to be $\Pi_L \gtrsim 0.8$.
In the case of photons produced in the cluster and $B_{\rm ext}<10^{-15} \, \rm G$, the final resulting photon-ALP conversion is less efficient, as it takes place inside the cluster and in the Milky Way only. In addition, Fig.~\ref{densProbHE} shows that a longer and more efficient photon-ALP conversion produces higher values of $\Pi_L$: this fact takes place by passing from the case of photon production in the cluster to that of photon emission in the blazar jet on one side and from the situation $B_{\rm ext}<10^{-15} \, \rm G$ to $B_{\rm ext}=1 \, \rm nG$ and $z=0.03$ and even more to $B_{\rm ext}=1 \, \rm nG$ and $z=0.4$ on the other. In addition, the effect of a high $B_{\rm ext}$ is to broaden $f_{\Pi}$, as discussed in the UV-X-ray band. This fact is amplified for $z=0.4$ with respect to $z=0.03$, since the extent of the extragalactic space, where photons can oscillate into ALPs, is larger in the former case.

\begin{figure*}
\centering
\includegraphics[width=0.96\textwidth]{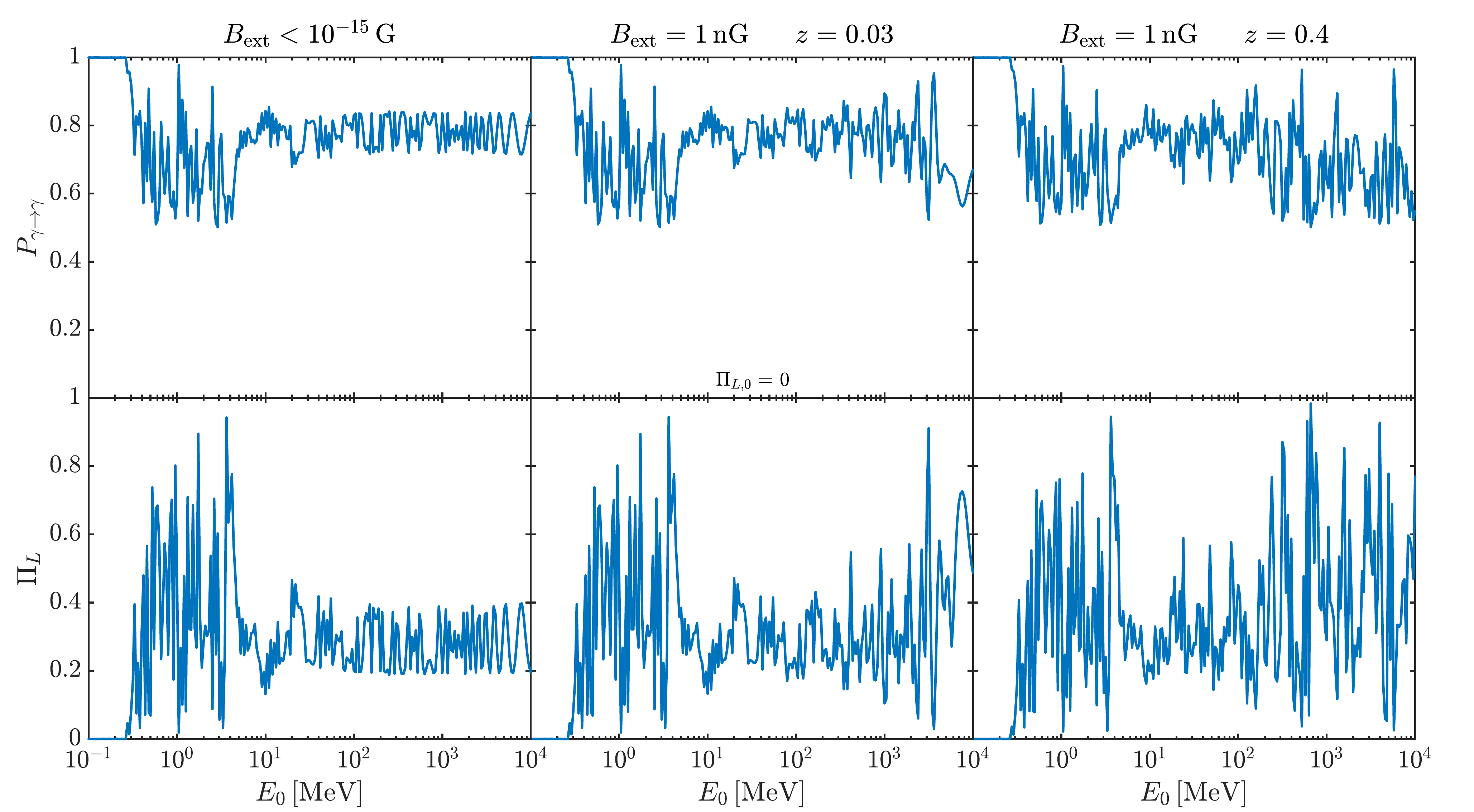}
\caption{\label{ProbPolMeV} Photon survival probability $P_{\gamma \to \gamma }$ (upper panels) and corresponding final photon degree of linear polarization $\Pi_L$ (lower panels) in the energy range $(10^{-1}-10^4) \, \rm MeV$ after propagation from the cluster, where photons are produced, up to us by taking $g_{a\gamma\gamma}=0.5 \times 10^{-11} \, \rm GeV^{-1}$, $m_a = 10^{-10} \, \rm eV$ and $n_{e,0}^{\rm clu} = 0.5 \times 10^{-2} \, \rm cm^{-3}$. The initial photon degree of linear polarization is $\Pi_{L,0}=0$. In the first column an extragalactic magnetic field $B_{\rm ext} < 10^{-15} \, \rm G$ is assumed. In the second column we take $B_{\rm ext} = 1 \, \rm nG$ and a redshift $z=0.03$. In the third column we consider $B_{\rm ext} = 1 \, \rm nG$ and $z=0.4$.}
\end{figure*}

\begin{figure*}
\centering
\includegraphics[width=0.96\textwidth]{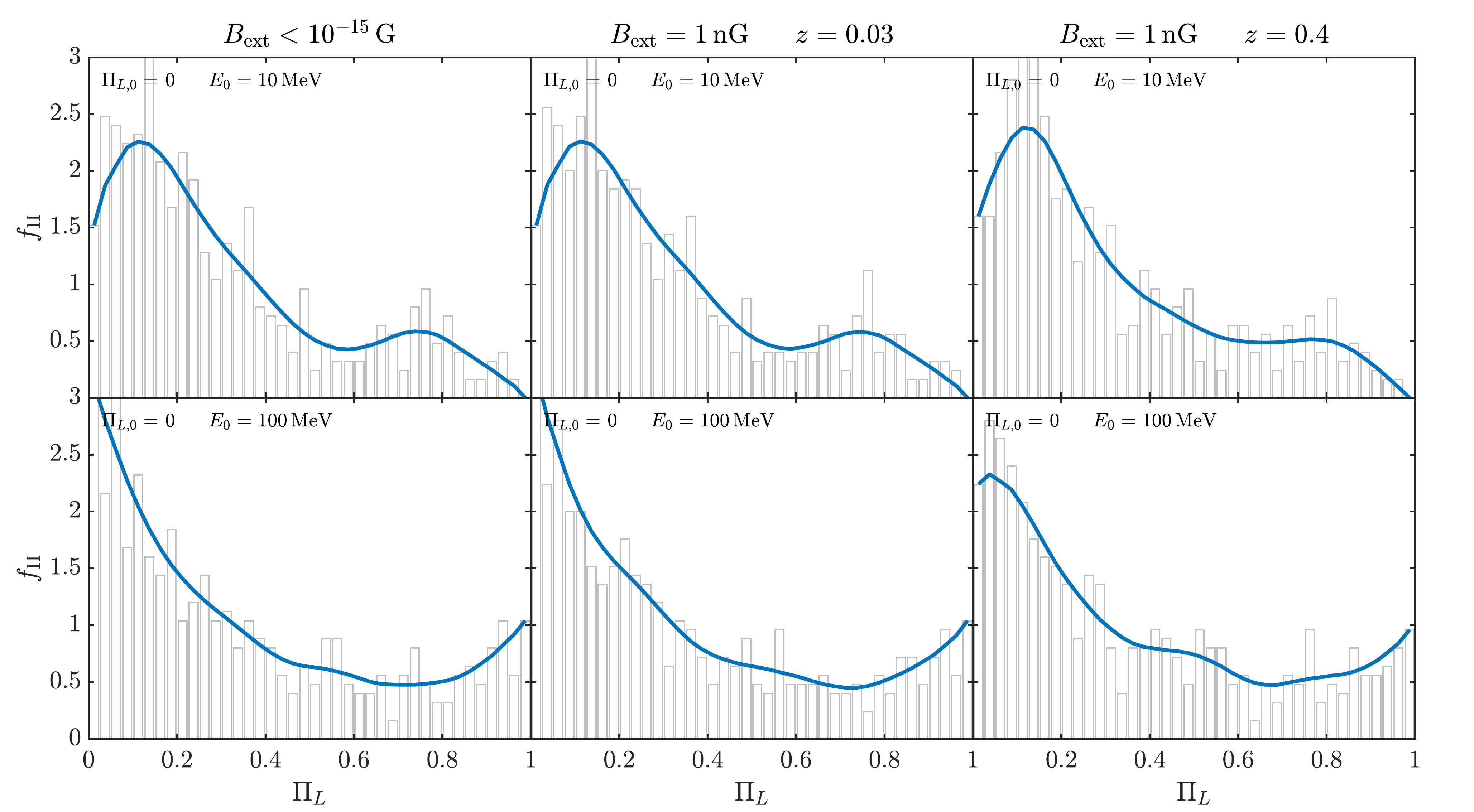}
\caption{\label{densProbMeV} Probability density function $f_{\Pi}$ arising from the plotted histogram for the final photon degree of linear polarization $\Pi_L$ at $10 \, \rm MeV$ (upper panels) and $100 \, \rm MeV$ (lower panels) by considering the system described in Fig.~\ref{ProbPolMeV}. The initial photon degree of linear polarization is $\Pi_{L,0}=0$. In the first column an extragalactic magnetic field $B_{\rm ext} < 10^{-15} \, \rm G$ is assumed. In the second column we take $B_{\rm ext} = 1 \, \rm nG$ and a redshift $z=0.03$. In the third column we consider $B_{\rm ext} = 1 \, \rm nG$ and $z=0.4$.}
\end{figure*}

\begin{figure*}
\centering
\includegraphics[width=0.96\textwidth]{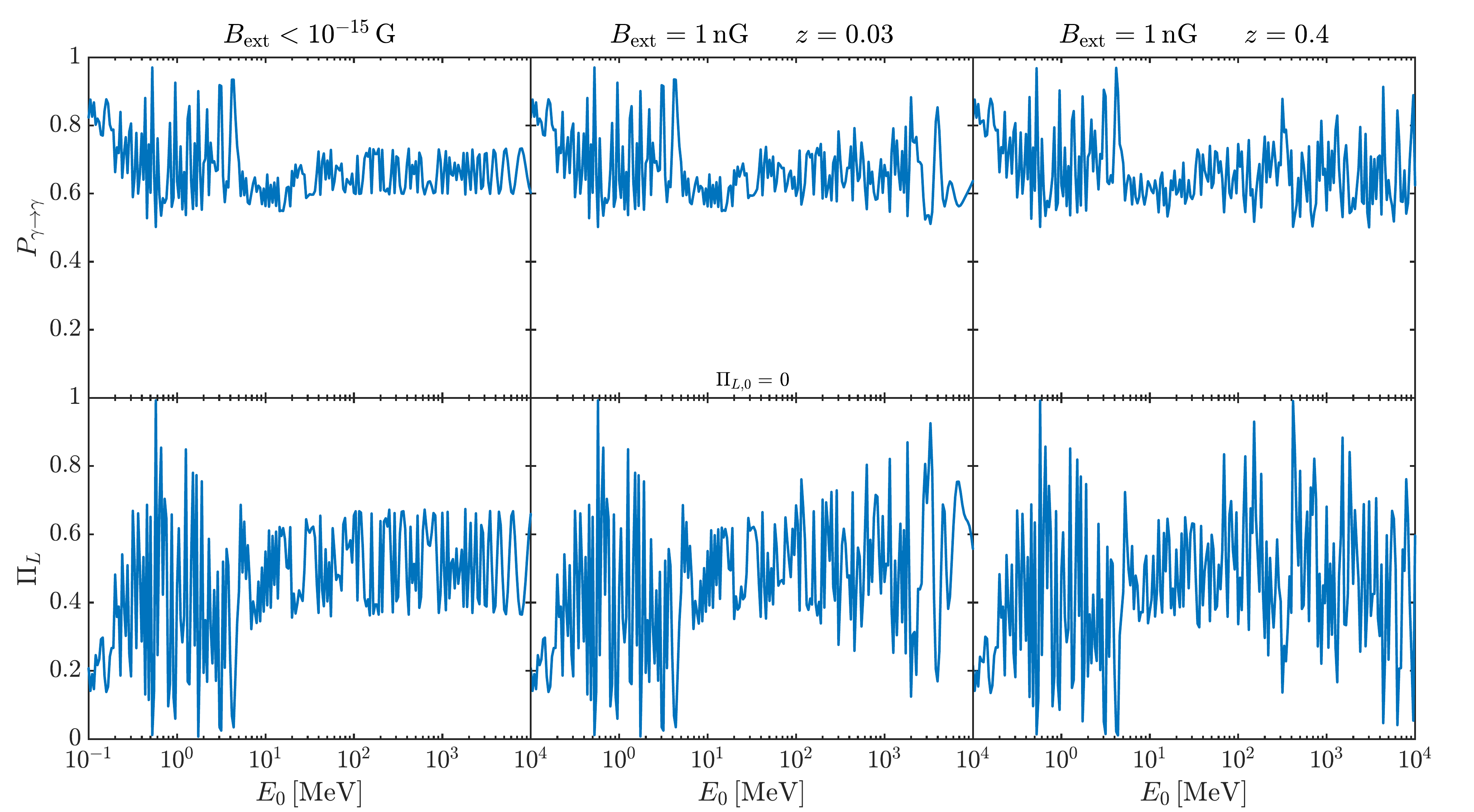}
\caption{\label{ProbPolMeVS} Same as Fig.~\ref{ProbPolMeV} but with also photon-ALP conversion within the blazar jet, where photons are produced. Thus, we accordingly take $n_{e,0}^{\rm clu} = 5 \times 10^{-2} \, \rm cm^{-3}$. The initial photon degree of linear polarization is $\Pi_{L,0}=0$.}
\end{figure*}

\begin{figure*}
\centering
\includegraphics[width=0.96\textwidth]{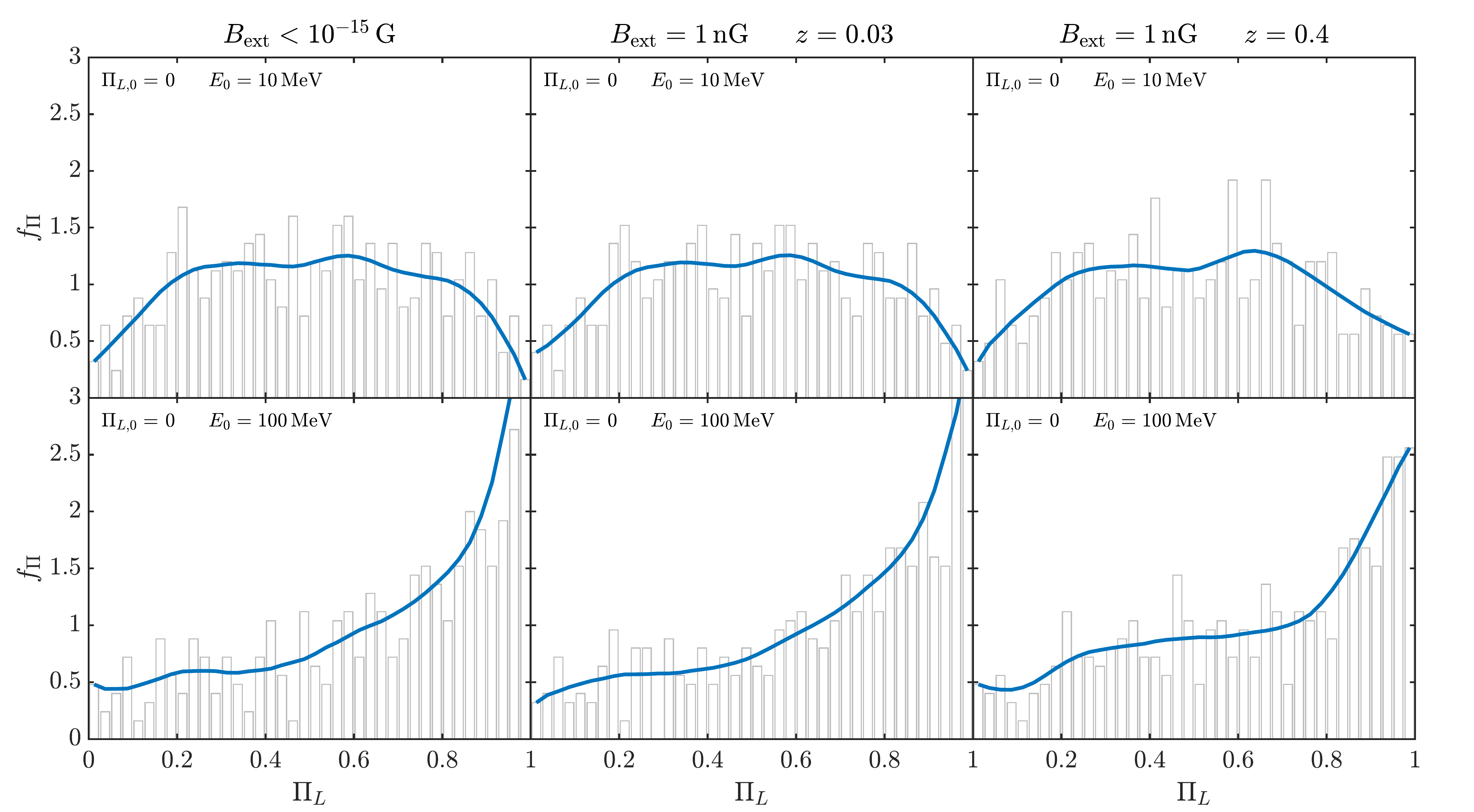}
\caption{\label{densProbMeVS} Same as Fig.~\ref{densProbMeV} but with also photon-ALP conversion within the blazar jet by considering the system described in Fig.~\ref{ProbPolMeVS}. The initial photon degree of linear polarization is $\Pi_{L,0}=0$.}
\end{figure*}

We now move to the case of an ALP with mass $m_a = 10^{-10} \, \rm eV$. In the present situation, the photon-ALP beam propagates in the weak mixing regime in all the energy band considered here ($10^{-1} \, {\rm MeV} - 10^4 \, \rm MeV$), as we can infer by the calculation of $E_L$ of Eq.~(\ref{EL}). Our results for the case of photons produced in the cluster are shown in Figs.~\ref{ProbPolMeV} and~\ref{densProbMeV}, while we report in Figs.~\ref{ProbPolMeVS} and~\ref{densProbMeVS} our findings concerning the alternate scenario of photons generated inside the jet of a blazar.

When photons are produced in the cluster, we take $n_{e,0}^{\rm clu}= 0.5 \times 10^{-2} \, \rm cm^{-3}$, which corresponds to a nCC galaxy cluster (see also Sec. IV.C). In Fig.~\ref{ProbPolMeV} we report $P_{\gamma \to \gamma}$ and the corresponding $\Pi_L$ for the different choices of $B_{\rm ext}$ and redshifts. From Fig.~\ref{ProbPolMeV}, we observe that the weak mixing regime extends for the entire energy range analyzed here ($10^{-1} \, {\rm MeV} - 10^4 \, \rm MeV$).
The very large extent of the energy range, where the weak mixing takes place, is due to the high variety of the properties of the media crossed by the photon-ALP beam (galaxy cluster, extragalactic space, Milky Way): as a result, $E_L$ greatly varies in the different zones. We note from the first row of Fig.~\ref{ProbPolMeV} that $P_{\gamma \to \gamma}$ never decreases down 0.5 as stated by item (ii) of Sec. III (see~\cite{galantiTheorems} for more details). The second row of Fig.~\ref{ProbPolMeV} shows that the final $\Pi_L$ turns out to be greatly modified with respect to $\Pi_{L,0}$ in almost all the energy band by the photon-ALP interaction. As $P_{\gamma \to \gamma}$ in the first row of Fig.~\ref{ProbPolMeV} confirms, we note a strong energy dependence of $\Pi_L$ (see the second row of Fig.~\ref{ProbPolMeV}), as the system lies in the weak mixing regime: the case of $B_{\rm ext}<10^{-15} \, \rm G$ is qualitatively similar to the ones of $B_{\rm ext}=1 \, \rm nG$ with $z=0.03$ and of $B_{\rm ext}=1 \, \rm nG$ with $z=0.4$ for $E \lesssim 100 \, \rm MeV$, while they differ for higher energies, where the photon-ALP interaction in the extragalactic space produces a sizable effect for $B_{\rm ext}=1 \, \rm nG$.

Because of the stochastic nature of the photon-ALP beam propagation process (see Sec. V.A) and in order to understand the impact of the photon-ALP interaction on $\Pi_L$, we plot the probability density function $f_{\Pi}$ of the final $\Pi_L$ associated to different realizations in Fig.~\ref{densProbMeV} for the two benchmark energies $E_0=10 \, \rm MeV$ and $E_0=100 \, \rm MeV$ (we recall that $P_{\gamma \to \gamma}$ and $\Pi_L$ in Fig.~\ref{ProbPolMeV} are associated to one realization of the propagation process). From Fig.~\ref{densProbMeV} we infer that the photon-ALP interaction produces a sizable variation of the initial $\Pi_{L,0}=0$ in all the cases. In addition, the most probable value of $\Pi_L$ is never $\Pi_L=0$ for $E_0=10 \, \rm MeV$, while it remains the most probable one for $E_0=100 \, \rm MeV$, when the photon-ALP interaction in the extragalactic space is not efficient enough ($B_{\rm ext} < 10^{-15} \, \rm G$ and $B_{\rm ext}=1 \, \rm nG$ with $z=0.03$).

We now move to the case of photons emitted at the blazar jet base, and we accordingly take $n_{e,0}^{\rm clu}= 5 \times 10^{-2} \, \rm cm^{-3}$ which corresponds to a CC galaxy cluster (see also Sec. IV.C). Our results about $P_{\gamma \to \gamma}$ and the corresponding $\Pi_L$ are reported in Fig.~\ref{ProbPolMeVS} for the different choices of $B_{\rm ext}$ and redshifts. What we have stated about the extent of the weak mixing regime in the case of photon production inside the cluster still holds true in the present case and similar conclusions about $P_{\gamma \to \gamma}$ and $\Pi_L$ can be achieved: $\Pi_L$ is greatly modified with respect to the initial value $\Pi_{L,0}$ (see above for more details). We just have to add that the photon-ALP interaction inside the magnetic field of the jet modifies in a sizable way the behavior of $P_{\gamma \to \gamma}$ (first row of Fig.~\ref{ProbPolMeVS}) and of the corresponding $\Pi_L$ (second row of Fig.~\ref{ProbPolMeVS}) for energies smaller than $\sim 0.5 \, \rm MeV$ with respect to the corresponding cases of photon production in the cluster. The reason for this modification is that the photon-ALP conversion is efficient inside the blazar jet also for $E_0 \lesssim 0.5 \, \rm MeV$, but the same is not true inside the galaxy cluster. 
 As the first row of Fig.~\ref{ProbPolMeVS} shows, we can check that $P_{\gamma \to \gamma}$ never decreases down 0.5 as assured by item (ii) of Sec. III (see~\cite{galantiTheorems} for more details). 

In Fig.~\ref{densProbMeVS} we report the probability density function $f_{\Pi}$ of the final $\Pi_L$ associated to several realizations of the propagation process (see also Sec. V.A for discussion about the stochastic behavior of the system) for the two benchmark energies $E_0=10 \, \rm MeV$ (upper panels) and $E_0=100 \, \rm MeV$ (lower panels). The behavior of $f_{\Pi}$ is almost independent on the value of $B_{\rm ext}$ and of the redshift apart from a small increase of the broadening of $f_{\Pi}$ as the redshift grows, for the same reasons discussed in the UV-X-ray band. For $E_0=10 \, \rm MeV$ (upper panels of Fig.~\ref{densProbMeVS}) $\Pi_L=0$ is never the most probable value for the final $\Pi_L$, while for $E_0=100 \, \rm MeV$ (lower panels of Fig.~\ref{densProbMeVS}) the most probable value for the final $\Pi_L$ turns out to be $\Pi_L \gtrsim 0.8$. From our findings about such strong polarization features, we can conclude that in the HE band, the case of photon production inside the blazar jet represents a better opportunity with respect to photon production inside the galaxy cluster, in order to search for ALP-induced effects on $\Pi_L$.
 


\subsection{Very-high-energy band}

First of all, we want to stress that, while for the UV-X-ray and the HE bands our findings can be tested by current and planned observatories~\cite{ixpe,extp,xcalibur,ngxp,xpp,cosi,eastrogam1,eastrogam2,amego}, our results about the photon polarization in the VHE range are nowadays purely theoretical. However, as it will be clear below, some important features about photon polarization linked to the photon-ALP interaction arise. Such features can be used to detect ALPs and/or constrain ALP parameters in case new techniques will hopefully be available in the future to measure photon polarization even in the VHE band.

\begin{figure*}
\centering
\includegraphics[width=0.67\textwidth]{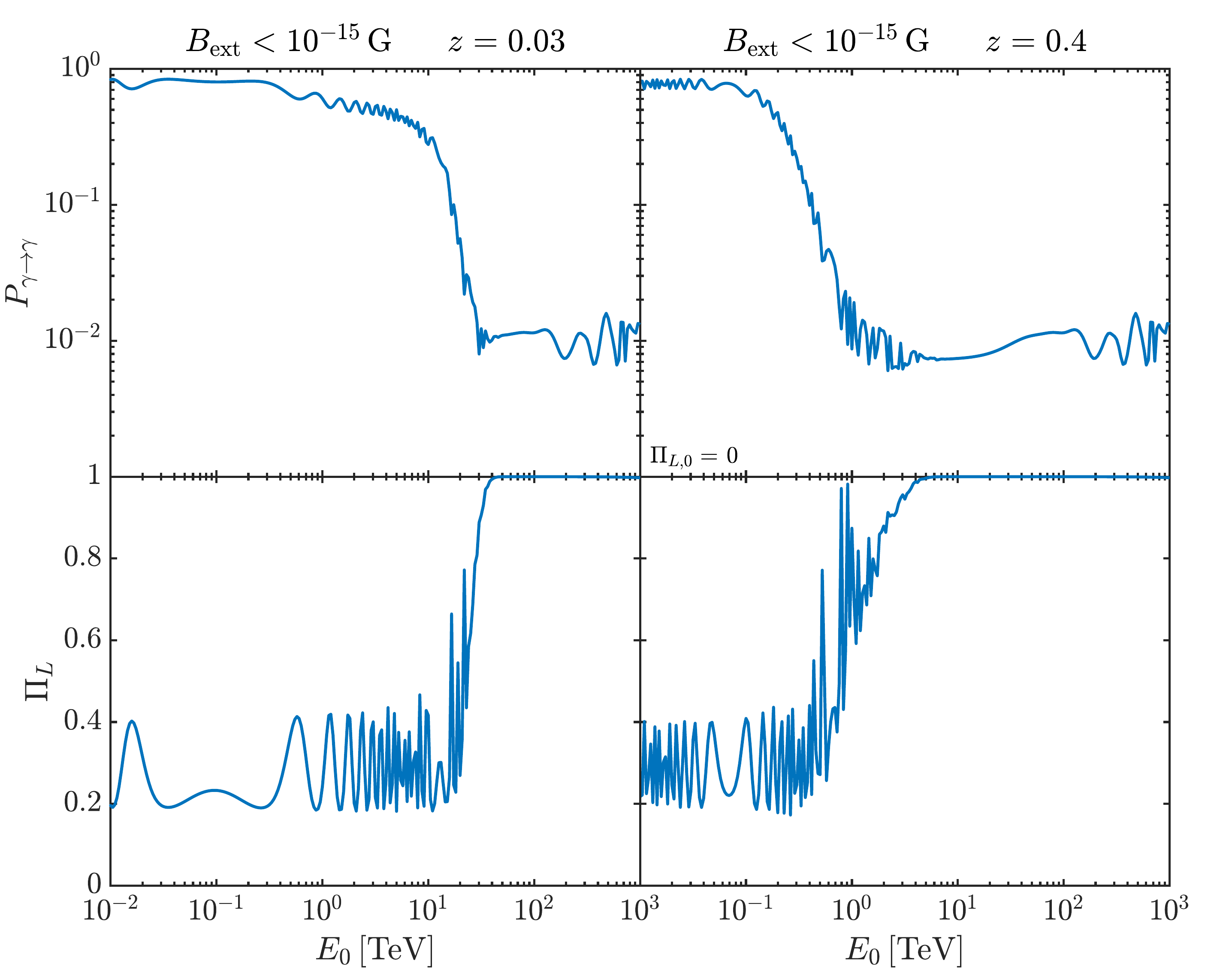}
\caption{\label{ProbPolVHEnoExt} Photon survival probability $P_{\gamma \to \gamma }$ (upper panels) and corresponding final photon degree of linear polarization $\Pi_L$ (lower panels) in the energy range $(10^{-2}-10^3) \, \rm TeV$ after propagation from the cluster, where photons are produced, up to us by taking $g_{a\gamma\gamma}=0.5 \times 10^{-11} \, \rm GeV^{-1}$, $m_a = 10^{-10} \, \rm eV$ and $n_{e,0}^{\rm clu} = 0.5 \times 10^{-2} \, \rm cm^{-3}$. An extragalactic magnetic field $B_{\rm ext} < 10^{-15} \, \rm G$ is assumed. The initial photon degree of linear polarization is $\Pi_{L,0}=0$. In the first column we take a redshift $z=0.03$, while in the second column we consider $z=0.4$.}
\end{figure*}

\begin{figure*}
\centering
\includegraphics[width=0.67\textwidth]{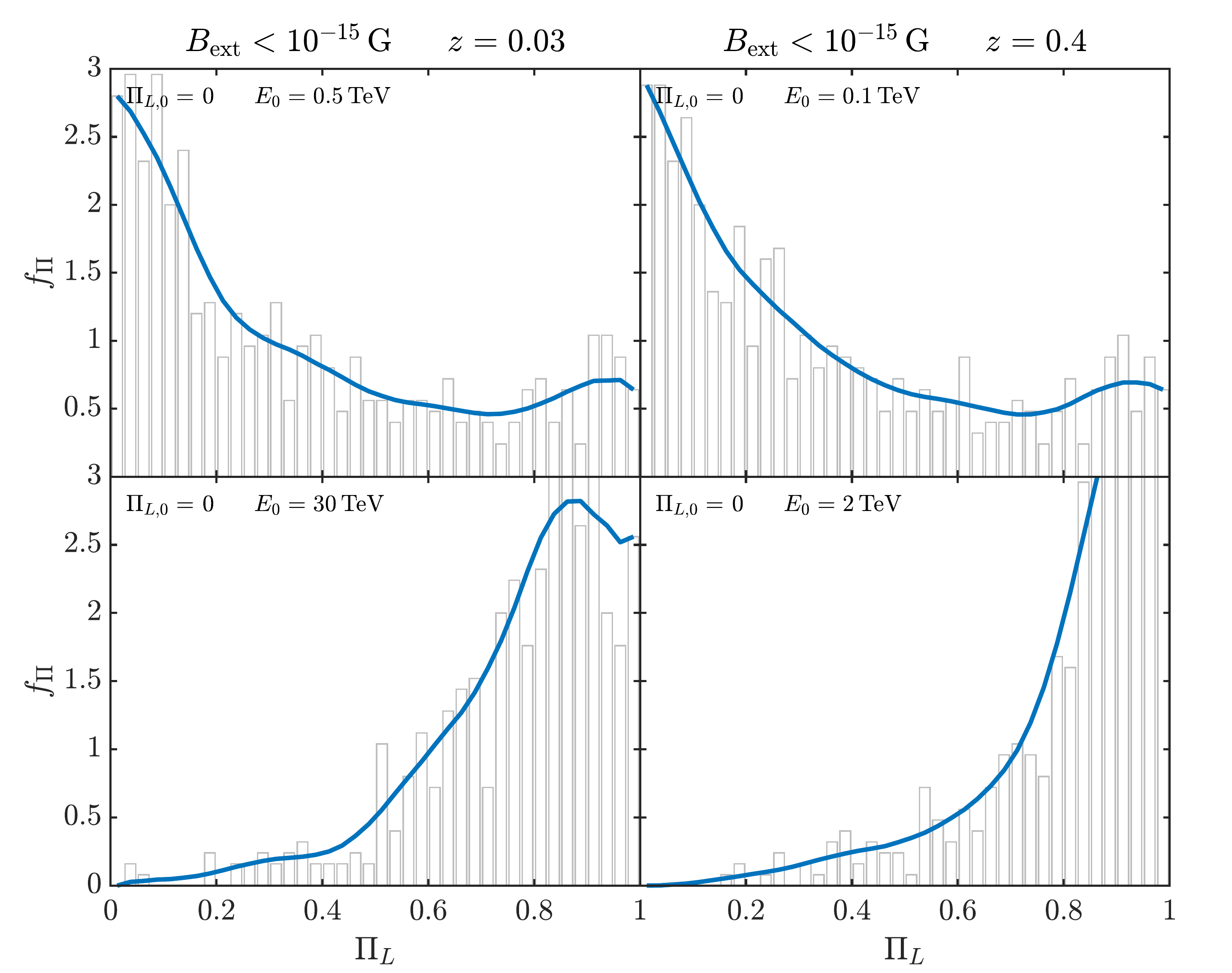}
\caption{\label{densProbVHEnoExt} Probability density function $f_{\Pi}$ arising from the plotted histogram for the final photon degree of linear polarization $\Pi_L$ at different energies (see subfigures) by considering the system described in Fig.~\ref{ProbPolVHEnoExt}. The initial photon degree of linear polarization is $\Pi_{L,0}=0$. In the first column we take a redshift $z=0.03$, while in the second column we consider $z=0.4$.}
\end{figure*}

\begin{figure*}
\centering
\includegraphics[width=0.67\textwidth]{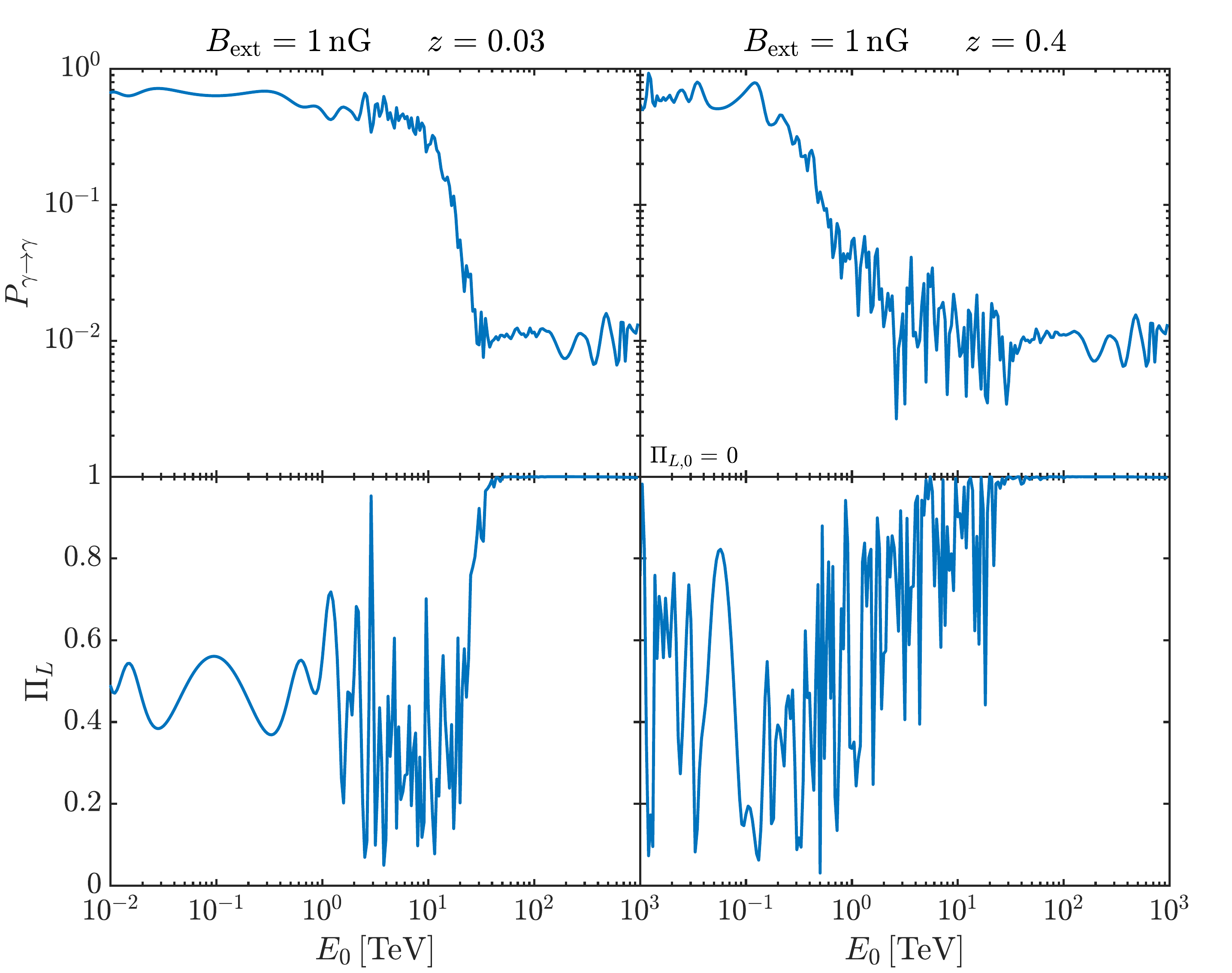}
\caption{\label{ProbPolVHEext} Same as Fig.~\ref{ProbPolVHEnoExt} but by considering an extragalactic magnetic field $B_{\rm ext} = 1 \, \rm nG$. The initial photon degree of linear polarization is $\Pi_{L,0}=0$.}
\end{figure*}

\begin{figure*}
\centering
\includegraphics[width=0.67\textwidth]{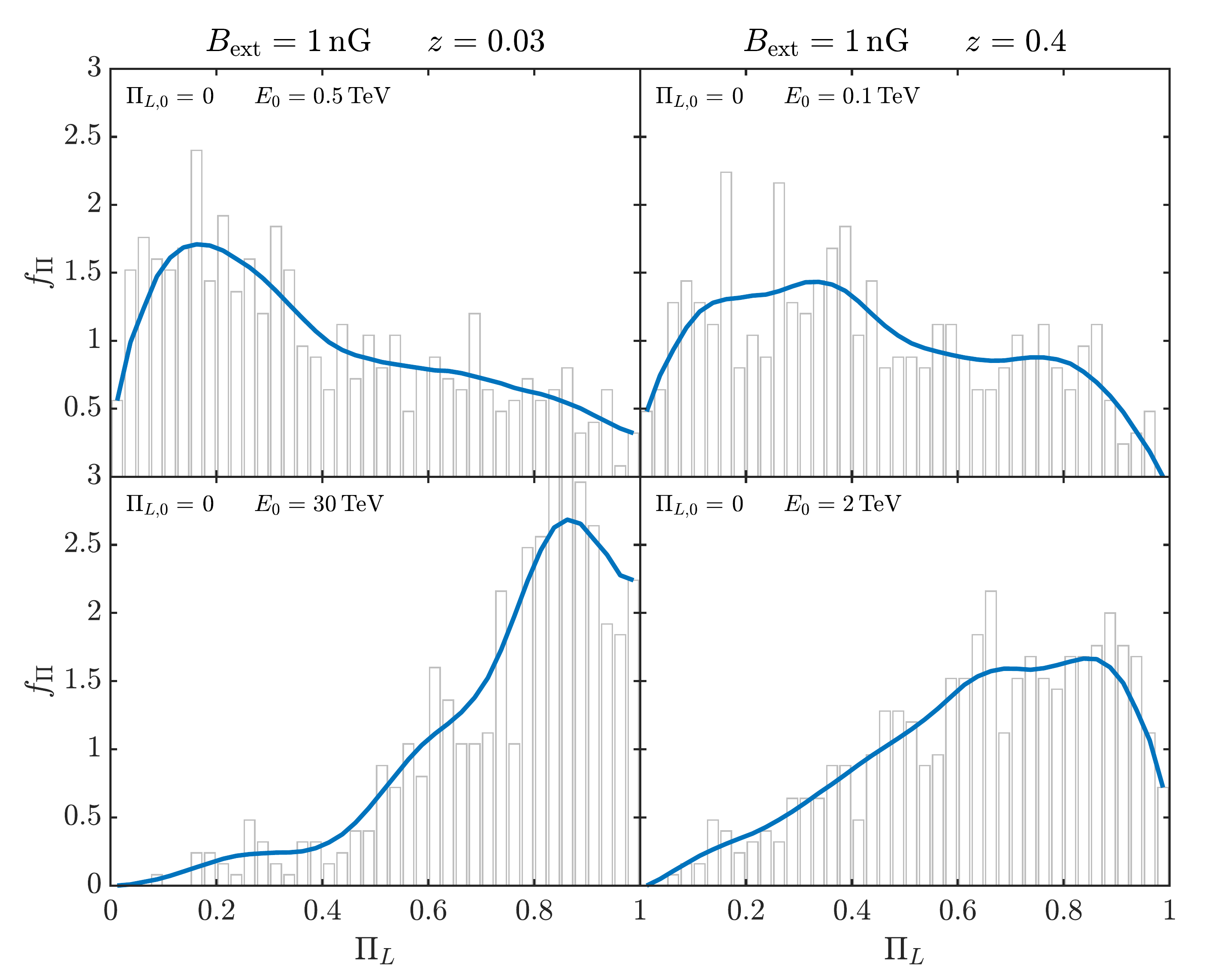}
\caption{\label{densProbVHEext} Same as Fig.~\ref{densProbVHEnoExt} but by considering the system described in Fig.~\ref{ProbPolVHEext}. The initial photon degree of linear polarization is $\Pi_{L,0}=0$.}
\end{figure*}

\begin{figure*}
\centering
\includegraphics[width=0.67\textwidth]{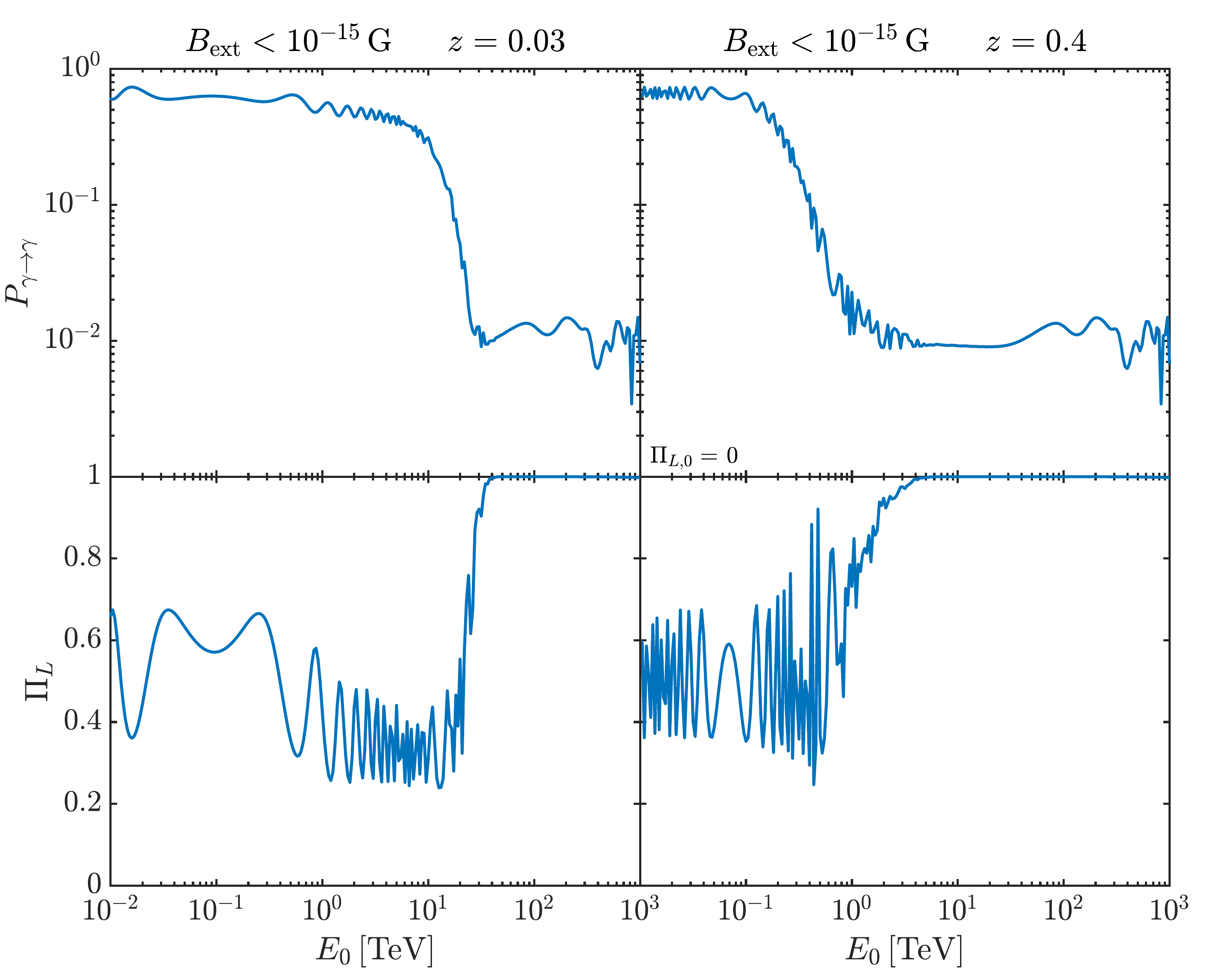}
\caption{\label{ProbPolVHEnoExtS} Same as Fig.~\ref{ProbPolVHEnoExt} but with also photon-ALP conversion within the blazar jet, where photons are produced. Thus, we accordingly take $n_{e,0}^{\rm clu} = 5 \times 10^{-2} \, \rm cm^{-3}$. The initial photon degree of linear polarization is $\Pi_{L,0}=0$.}
\end{figure*}

\begin{figure*}
\centering
\includegraphics[width=0.67\textwidth]{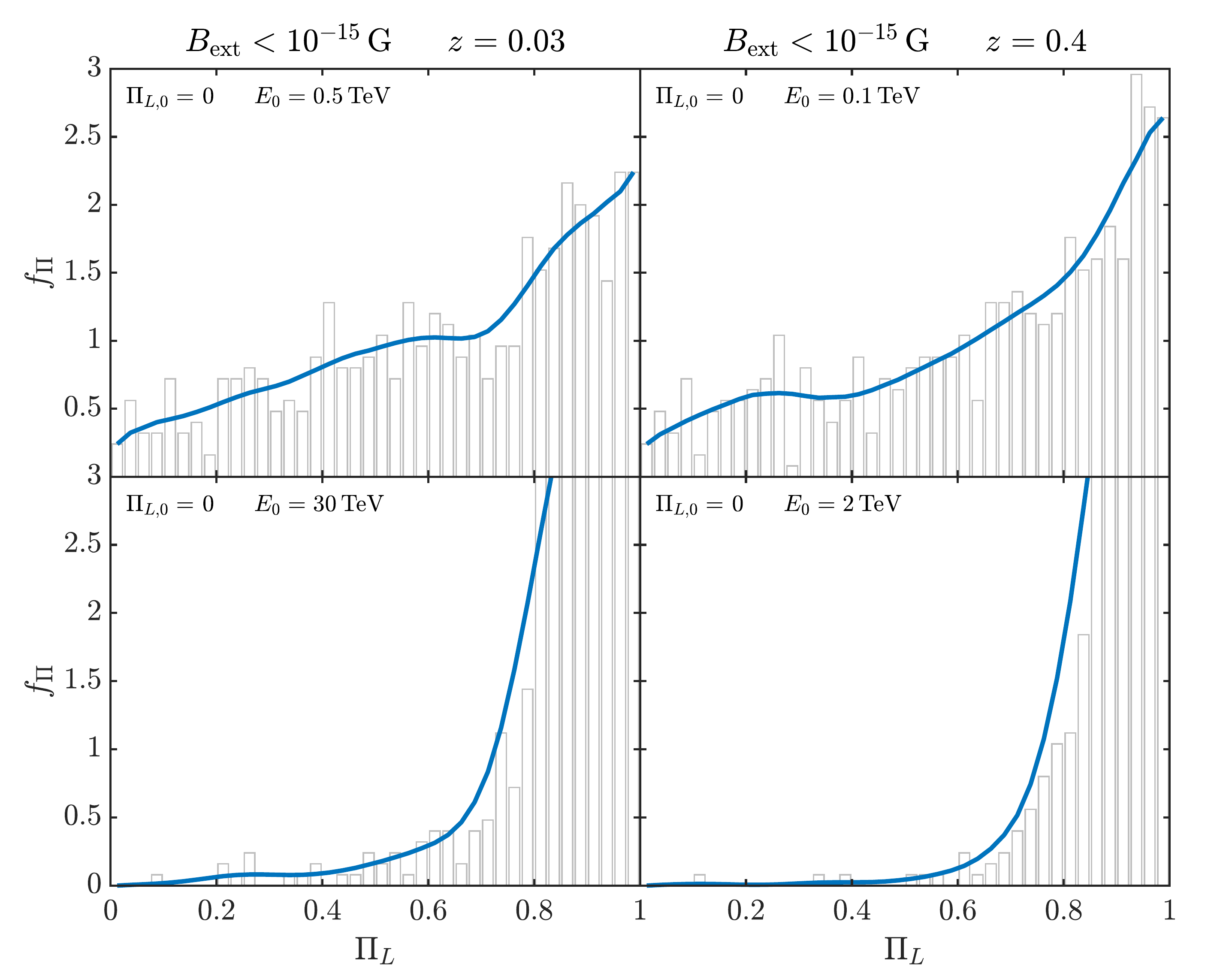}
\caption{\label{densProbVHEnoExtS} Same as Fig.~\ref{densProbVHEnoExt} but with also photon-ALP conversion within the blazar jet by considering the system described in Fig.~\ref{ProbPolVHEnoExtS}. The initial photon degree of linear polarization is $\Pi_{L,0}=0$.}
\end{figure*}

\begin{figure*}
\centering
\includegraphics[width=0.67\textwidth]{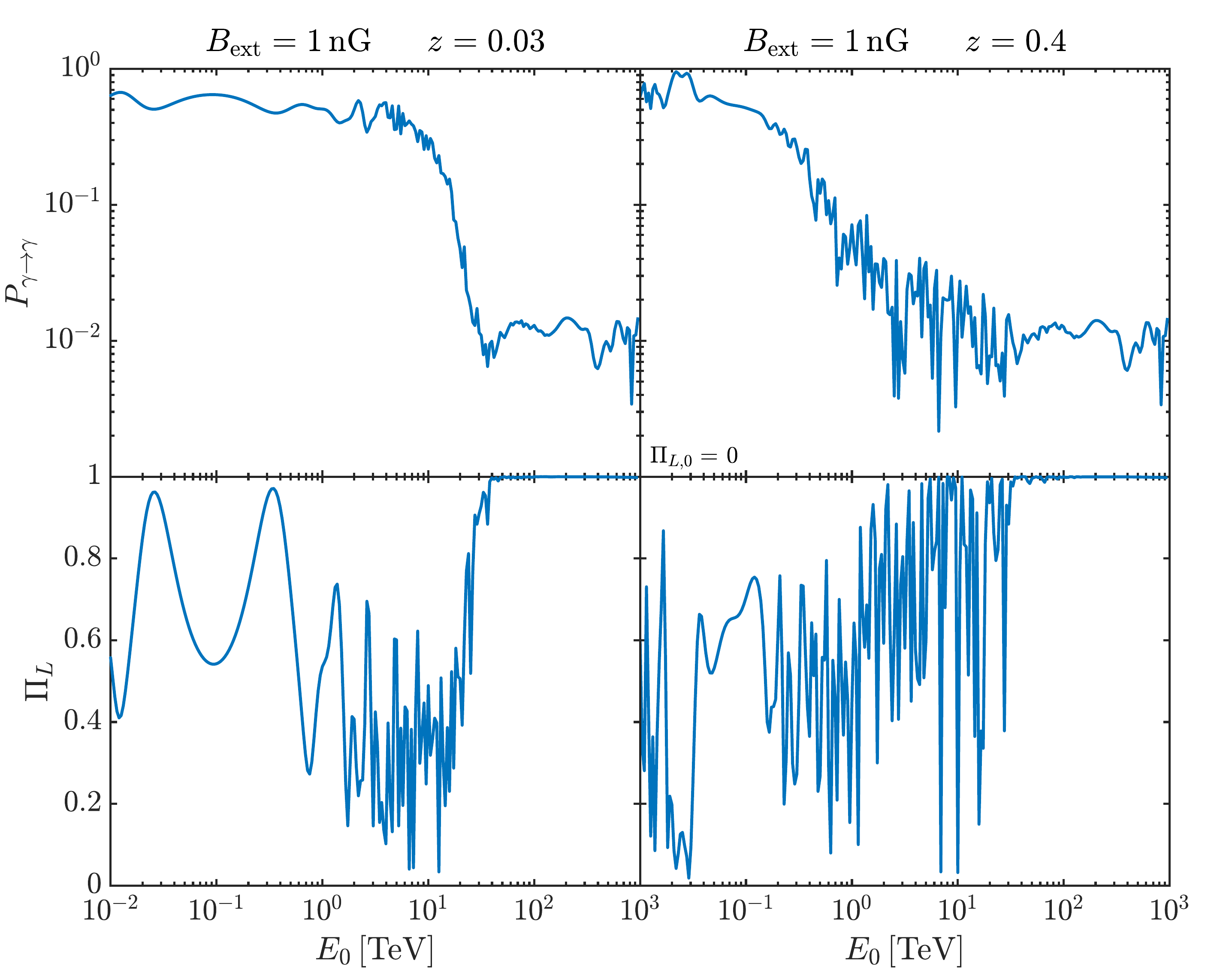}
\caption{\label{ProbPolVHEextS} Same as Fig.~\ref{ProbPolVHEnoExt} but with also photon-ALP conversion within the blazar jet, where photons are produced. Thus, we accordingly take $n_{e,0}^{\rm clu} = 5 \times 10^{-2} \, \rm cm^{-3}$. Here, we consider $B_{\rm ext} = 1 \, \rm nG$. The initial photon degree of linear polarization is $\Pi_{L,0}=0$.}
\end{figure*}

\begin{figure*}
\centering
\includegraphics[width=0.67\textwidth]{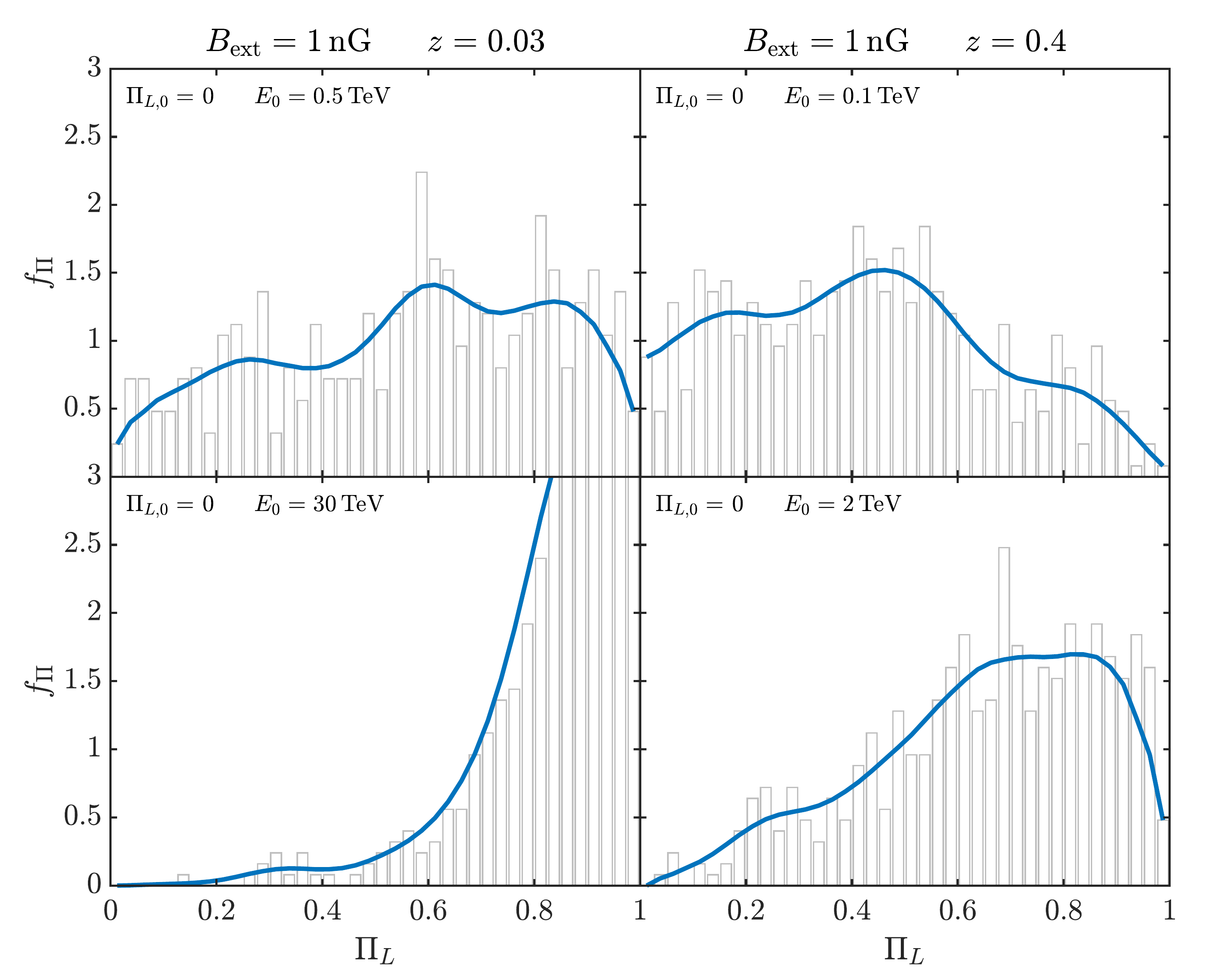}
\caption{\label{densProbVHEextS} Same as Fig.~\ref{densProbVHEnoExt} but with also photon-ALP conversion within the blazar jet by considering the system described in Fig.~\ref{ProbPolVHEextS}. The initial photon degree of linear polarization is $\Pi_{L,0}=0$.}
\end{figure*}

The calculation of $E_H$ from Eq.~(\ref{EH}) is made complicated by the fact that in the energy range $(10^{-2}-10^3) \, \rm TeV$ VHE photons are absorbed because of their interaction with the EBL photons producing an $e^+e^-$ pair through the process $\gamma \gamma \to e^+ e^-$. This process limits the $\gamma$-ray horizon more and more as the observed photon energy $E_0$ grows~\cite{dgr2013}. Nevertheless, by inspection of $P_{\gamma \to \gamma}$ in the following figures we infer that the photon-ALP system is never in the strong mixing regime in almost all the energy band considered here and we can observe that both $P_{\gamma \to \gamma}$ and the corresponding $\Pi_L$ become energy dependent.  In addition, because of the $\gamma \gamma$ absorption ${\cal U}_{\rm ext}$ is no more unitary and in the case of both $B_{\rm ext}<10^{-15} \, \rm G$ and $B_{\rm ext}=1 \, \rm nG$ the photon-ALP system is sensible to the distance traveled in the extragalactic space, so that in both the situations we consider the two redshifts $z=0.03$ and $z=0.4$. In particular, our results for the case of photons produced in the cluster are plotted in Figs.~\ref{ProbPolVHEnoExt} and~\ref{densProbVHEnoExt} for $B_{\rm ext}<10^{-15} \, \rm G$, and in Figs.~\ref{ProbPolVHEext} and~\ref{densProbVHEext} for $B_{\rm ext}=1 \, \rm nG$. Furthermore, we plot our findings for the case of photons generated in the blazar jet in Figs.~\ref{ProbPolVHEnoExtS} and~\ref{densProbVHEnoExtS} for $B_{\rm ext}<10^{-15} \, \rm G$, and in Figs.~\ref{ProbPolVHEextS} and~\ref{densProbVHEextS} for $B_{\rm ext}=1 \, \rm nG$. Similarly to the UV-X-ray and HE band, when photons are emitted in the cluster, we consider $n_{e,0}^{\rm clu}= 0.5 \times 10^{-2} \, \rm cm^{-3}$, which corresponds to a nCC galaxy cluster, while if photons are produced in the blazar jet, we correspondingly take a CC galaxy cluster with $n_{e,0}^{\rm clu}= 5 \times 10^{-2} \, \rm cm^{-3}$ (see also Sec. IV.C). Throughout this Section we consider an ALP with mass $m_a = 10^{-10} \, \rm eV$. The case $m_a \lesssim 10^{-14} \, \rm eV$ slightly differs from the previous one for $E_0 \lesssim 50 \, \rm GeV$ only, where the system is in the strong mixing regime and $P_{\gamma \to \gamma}$ and $\Pi_L$ are energy independent. Therefore, what we have found in the HE band in the case of $m_a \lesssim 10^{-14} \, \rm eV$ perfectly extends here in the VHE range for $E_0 \lesssim 50 \, \rm GeV$ and $m_a \lesssim 10^{-14} \, \rm eV$. Hereafter, we thus concentrate on the case $m_a = 10^{-10} \, \rm eV$.

What we have discussed in the HE band about the initial photon degree of linear polarization $\Pi_{L,0}$ still holds true for the VHE range, so that we assume photons as initially unpolarized with initial $\Pi_{L,0}=0$ in both the cases of photons produced either inside the cluster or in the blazar jet.

In all the figures concerning $P_{\gamma \to \gamma}$ we observe that $P_{\gamma \to \gamma}$ starts to decrease in a sizable way because of the EBL $\gamma \gamma$ absorption for $E_0 \gtrsim 3 \, \rm TeV$ at $z=0.03$ and for $E_0 \gtrsim 200 \, \rm GeV$ at $z=0.4$. Correspondingly, we observe an increase of $\Pi_L$ up to the limit value $\Pi_L=1$ -- with photons totally polarized -- for $E_0 \gtrsim 30 \, \rm TeV$ in the case $z=0.03$ and for $E_0 \gtrsim 5 \, \rm TeV$ in the case $z=0.4$: this fact takes place where the associated $P_{\gamma \to \gamma} \lesssim 10^{-2}$. The reason for this behavior concerning $\Pi_L$ lies in the growing $\gamma \gamma$ absorption due to the EBL as $E_0$ increases. For energies where absorption is not dramatic -- $E_0 \lesssim 3 \, \rm TeV$ at $z=0.03$ and $E_0 \lesssim 200 \, \rm GeV$ at $z=0.4$ -- what occurs in the VHE range is totally similar to the UV-X-ray and HE band. In fact, we observe that, when absorption is not too high, $\Pi_L$ moderately increases above the initial value $\Pi_{L,0}=0$ showing an energy-dependent behavior, since the photon-ALP system is in the weak mixing regime. In the case of high absorption, instead, what takes place can be visualized as follows. When photons are produced (either in the cluster or in the blazar jet), they partially convert into ALPs while crossing the magnetized media close to the source (blazar jet magnetic field ${\bf B}^{\rm jet}$ and/or galaxy cluster magnetic field ${\bf B}^{\rm clu}$), so that before the photon-ALP beam propagates inside the extragalactic space is made of both photons and ALPs. Conversion inside the extragalactic space may take place ($B_{\rm ext} = 1 \, \rm nG$) or not ($B_{\rm ext} < 10^{-15} \, \rm G$), but it is in any case not efficient as the system lies in the weak mixing regime because of the photon dispersion on the CMB (see also~\cite{grExt}). Therefore, while photons are almost totally absorbed, a sizable amount of ALPs survives up to the Milky Way, where ALPs can reconvert back to photons inside the magnetic field of the Milky Way ${\bf B}_{\rm MW}$. Since what is efficient for the photon-ALP conversion inside the Milky Way is the coherent part of ${\bf B}_{\rm MW}$, photons reconverted back from ALPs inside the Milky Way are fully polarized. This behavior is valid at all energies, in the UV-X-ray, HE and VHE band, but in case of no/low absorption (UV-X-ray and HE band) it is hidden by the presence of the photons that oscillate into ALPs in other regions outside the Milky Way. Instead, when absorption is very high -- i.e. in the VHE band -- almost all photons apart from those reconverted back from ALPs in the Milky Way are absorbed in the extragalactic space because of their interaction with the EBL. This is the reason why the final $\Pi_L$ grows towards the limit value $\Pi_L=1$ as the photon energy grows, in the same energy range where $\gamma\gamma$ absorption due to the EBL grows as well.

We can observe that Figs.~\ref{ProbPolVHEnoExt} and~\ref{ProbPolVHEext} showing $P_{\gamma \to \gamma}$ and $\Pi_L$ for photon production in the cluster for the cases of $B_{\rm ext} < 10^{-15} \, \rm G$ and $B_{\rm ext} = 1 \, \rm nG$, respectively and Figs.~\ref{ProbPolVHEnoExtS} and~\ref{ProbPolVHEextS} exhibiting $P_{\gamma \to \gamma}$ and $\Pi_L$ for photon emission in the blazar jet for the cases of $B_{\rm ext} < 10^{-15} \, \rm G$ and $B_{\rm ext} = 1 \, \rm nG$, respectively are all qualitatively similar and described by the behavior discussed above. We note that, when $B_{\rm ext} = 1 \, \rm nG$, the photon-ALP conversion in the extragalactic space produces more oscillations in $P_{\gamma \to \gamma}$ and $\Pi_L$ with respect to the energy if compared to the case $B_{\rm ext} < 10^{-15} \, \rm G$ when $E_0 \lesssim 50 \, \rm TeV$. Above this energy the effect of the photon dispersion on the CMB is so strong that the photon-ALP interaction in the extragalactic space is completely inefficient (see also~\cite{grExt}) and we do not observe any difference between the cases $B_{\rm ext} < 10^{-15} \, \rm G$ and $B_{\rm ext} = 1 \, \rm nG$.

In order to infer the statistical properties of the photon-ALP system and the robustness of our results about the final $\Pi_L$, we analyze the probability density function $f_{\Pi}$ of $\Pi_L$ associated to several realizations of the photon-ALP beam propagation process. Thus, for different energies, we plot in Figs.~\ref{densProbVHEnoExt} and~\ref{densProbVHEext} $f_{\Pi}$ for photon production in the cluster in the cases of $B_{\rm ext} < 10^{-15} \, \rm G$ and $B_{\rm ext} = 1 \, \rm nG$, respectively, and we report in Figs.~\ref{densProbVHEnoExtS} and~\ref{densProbVHEextS} $f_{\Pi}$ for photon generation in the blazar jet in the cases of $B_{\rm ext} < 10^{-15} \, \rm G$ and $B_{\rm ext} = 1 \, \rm nG$, respectively. In particular, in all the above-mentioned figures we consider $E_0 = 500 \, \rm GeV$ and $E_0 = 30 \, \rm TeV$ when $z=0.03$ and $E_0 = 100 \, \rm GeV$ and $E_0 = 2 \, \rm TeV$ when $z=0.4$. We take lower energies when the redshift grows since the EBL $\gamma \gamma$ absorption increases with the enhancement of both energy and distance, so that the behavior of $f_{\Pi}$ at the two redshifts becomes comparable for the considered energies. Correspondingly, in all the figures about $f_{\Pi}$ we have low absorption in both the cases of $E_0 = 500 \, \rm GeV$ with $z=0.03$ and $E_0 = 100 \, \rm GeV$ with $z=0.4$: in the present situation the photon-ALP conversion broadens and increases the initial $\Pi_{L,0}$ so that the final $\Pi_L= 0$ is never the most probable value in all the figures apart from the case of photon production in the cluster and $B_{\rm ext} < 10^{-15} \, \rm G$. Instead, in both the situations of $E_0 = 30 \, \rm TeV$ with $z=0.03$ and $E_0 = 2 \, \rm TeV$ with $z=0.4$, the EBL absorption is very high, so that the greatest part of the detectable photons are those reconverted back from ALPs inside the Milky Way, as discussed above. In fact, in the present case the most probable value for the final $\Pi_L$ becomes $\Pi_L \gtrsim 0.8$. Obviously, by increasing $E_0$ also the final $\Pi_L$ grows up to its limit value $\Pi_L=1$. In the presence of low absorption the most probable value for the final $\Pi_L$ is higher in the case of photon generation inside the blazar jet with respect to the case of photon production in the cluster.  As already observed in the UV-X-ray and HE band, the effect of $B_{\rm ext} = 1 \, \rm nG$ is to broaden the value of $\Pi_L$, for the same reasons discussed in the previous Subsections. The latter fact is more evident in the case $z = 0.4$ due to the larger distance covered by the photon-ALP beam in the extragalactic space, as shown by the comparison of the lower-right panel of Fig.~\ref{densProbVHEnoExt} with the lower-right panel of Fig.~\ref{densProbVHEext} and of the lower-right panel of Fig.~\ref{densProbVHEnoExtS} with the lower-right panel of Fig.~\ref{densProbVHEextS}.


\section{Discussion}

In the previous Sections we have shown that the photon-ALP interaction produces several features on the final photon polarization, which may be detectable by current and future observatories. ALP-induced polarization effects are mainly produced by the photon-ALP interaction inside the blazar jet and/or in the cluster, while the contribution of other regions is less effective (see previous Sections). A detection of signals in contrast with conventional physics expectations, as a final photon degree of linear polarization $\Pi_L > 0$ for photons coming from galaxy clusters, would represent a hint for new physics in terms of ALPs .

We now cursorily examine the real possibility of observing the above discussed characteristics but we defer a deeper analysis in this respect in forthcoming papers.

Concerning the real detectability of the above described features in the X-ray band, we must be aware that polarization measurements are more difficult with respect to the flux ones, so that a lower energy resolution is likely: we empirically consider an energy resolution worse by a factor $4-5$ for polarization observations with respect to flux surveys. Hence, by assuming the energy resolution of current flux-measuring X-ray observatories, we expect that $15-20$ energy bins per decade can be resolved by polarimeters in the X-ray band~\cite{swift}.  
Therefore, we expect observatories like IXPE~\cite{ixpe}, eXTP~\cite{extp}, XL-Calibur~\cite{xcalibur}, NGXP~\cite{ngxp} and XPP~\cite{xpp} to possess enough energy resolution to be able to detect ALP effects on photon polarization and its features especially for $E_0 \gtrsim 1 \, \rm keV$. 

In the HE range, spectral and polarimetric measurements are expected to possess a similar energy resolution, as they derive from the same data. Correspondingly, we conservatively assume an energy resolution of $8 - 10$ bins per decade~\cite{eastrogam1,eastrogam2,amego}. Therefore, observatories like COSI~\cite{cosi}, e-ASTROGAM~\cite{eastrogam1,eastrogam2} and AMEGO~\cite{amego} are expected to be able to detect the ALP-induced modifications to photon polarization reported above.

Concerning photons emitted at the blazar jet base, we have to take into account also the limited spatial resolution of polarimeters: the instruments are unable to discriminate among photons coming from the different zones inside the transverse section of the blazar jet. As a result, polarization features could in principle be attenuated, since photons, experiencing different orientations of the magnetic field in the jet ${\bf B}^{\rm jet}$, are all collected together and photon polarization is thus averaged over the whole jet transverse section.

\begin{figure*}
\centering
\includegraphics[width=0.7\textwidth]{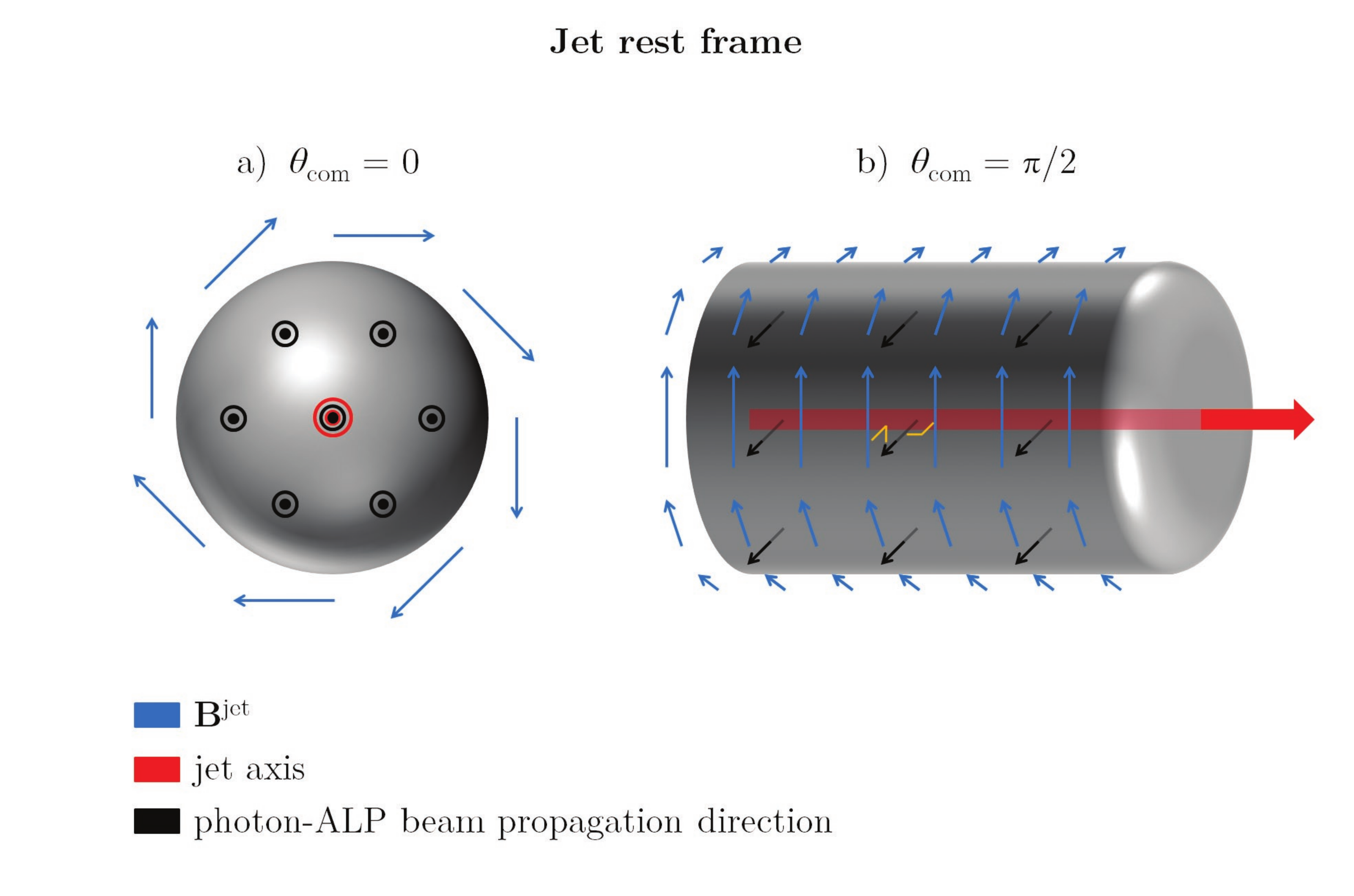}
\caption{\label{imageJet} Picture of the two extreme cases for the photon-ALP beam propagation in the jet rest frame. a) Left panel: perfect alignment between the photon-ALP beam propagation direction and the jet axis. 
b) Right panel: maximal misalignment with the photon-ALP beam propagating in the orthogonal direction with respect to the jet axis. 
}  
\end{figure*}

The behavior of the photon-ALP conversion inside the jet depends on the line of sight, as the photon-ALP beam experiences various ${\bf B}^{\rm jet}$ configurations (only the toroidal component  of ${\bf B}^{\rm jet}$ is relevant, see also Sec. IV.A), while propagating at different angles with respect to the jet axis. We call $\theta_{\rm com}$ the angle between the jet axis and the photon-ALP beam propagation direction in the rest frame of the jet, while $\theta_{\rm fix}$ represents the same angle but as seen in the fixed external frame. In Fig.~\ref{imageJet} we show, {\it in the jet rest frame}, the geometry of the two extreme cases: (i) perfect alignment between the photon-ALP beam propagation direction and the jet axis (see case a) in Fig.~\ref{imageJet}) with therefore $\theta_{\rm com}=\theta_{\rm fix}=0$; (ii) the photon-ALP beam orthogonally propagating with respect to the jet axis (see case b) in Fig.~\ref{imageJet}) with thus $\theta_{\rm com}= \pi/2$, resulting in $\theta_{\rm fix} \simeq 1/\gamma$ because 
of the aberration induced by the Lorentz factor $\gamma$.

In the case a) of Fig.~\ref{imageJet}, the photon-ALP beam experiences a ${\bf B}^{\rm jet}$ which is equally oriented in whatever direction inside the jet transverse section. We have maximal photon-ALP conversion since ${\bf B}^{\rm jet}$ is orthogonal to the photon-ALP beam momentum but the beam sees ${\bf B}^{\rm jet}$ with minimal symmetry: therefore, by collecting all photons together, ALP-induced polarization features are expected to be attenuated.

In the case b) of Fig.~\ref{imageJet}, instead, the transverse component of ${\bf B}^{\rm jet}$ (which is the only relevant for the photon-ALP conversion) is coherent and it turns out to be maximal in correspondence with the plane defined by the photon-ALP beam propagation direction and the jet axis. The region around this plane represents the zone where the maximal amount of photons reaching us are produced. Above and below this plane the transverse component of ${\bf B}^{\rm jet}$ decreases together with a smaller photon emission, since photons are produced from a smaller volume. In the present situation -- case b) of Fig.~\ref{imageJet} -- we have less photon-ALP conversion as the transverse component of ${\bf B}^{\rm jet}$ decreases far from the previously defined plane but maximal ${\bf B}^{\rm jet}$ symmetry: therefore, by collecting all photons together, ALP-induced polarization features are expected to be retained.

The case of a generic misalignment between the photon-ALP beam propagation direction and the jet axis is intermediate between case a) and b) of Fig.~\ref{imageJet}. 

\begin{figure*}
\centering
\includegraphics[width=0.67\textwidth]{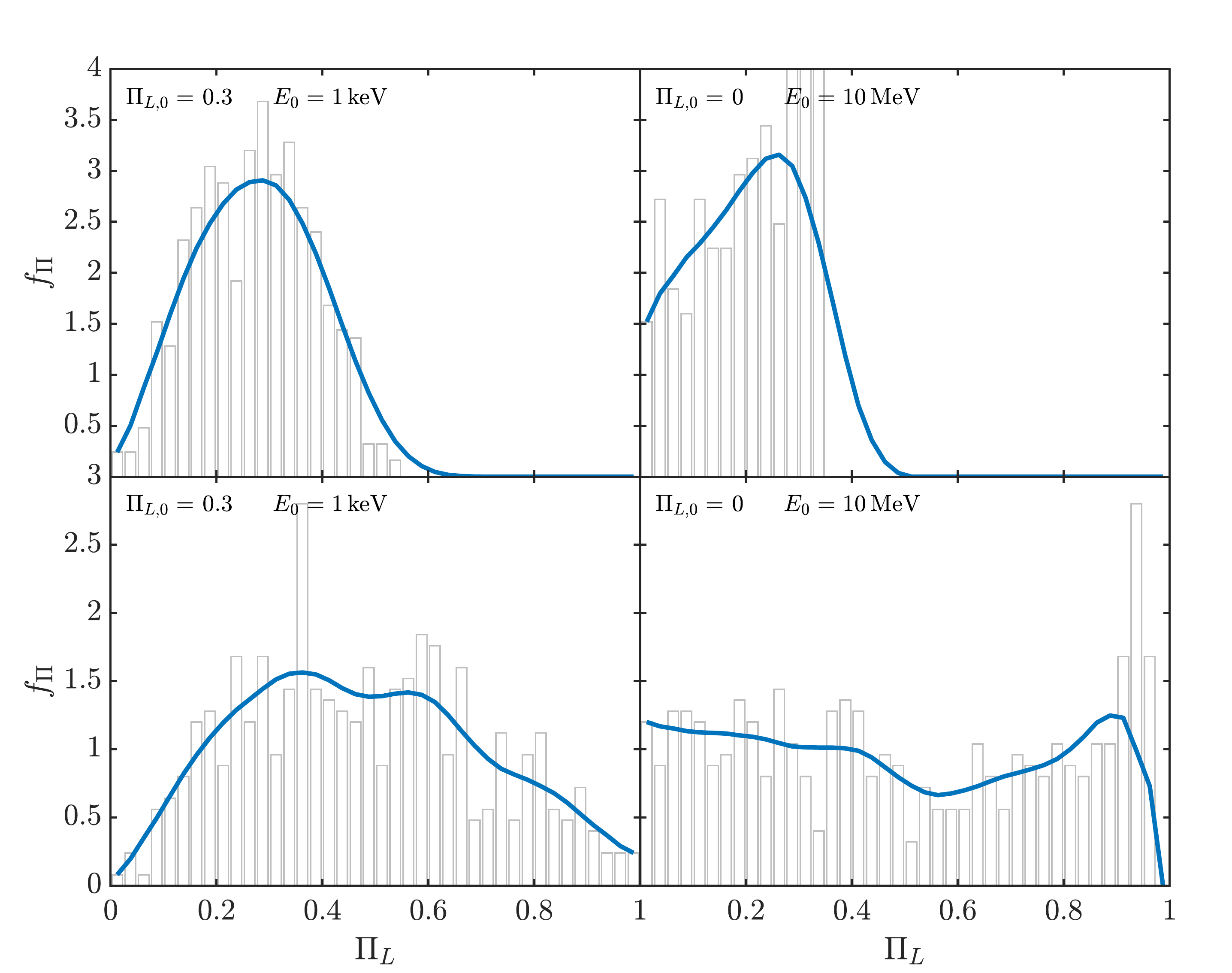}
\caption{\label{densProbBlazObserv} Probability density function $f_{\Pi}$ arising from the plotted histogram for the final photon degree of linear polarization $\Pi_L$ at $1 \, \rm keV$ (left panels) and $10 \, \rm MeV$ (right panels) after propagation from the blazar jet, where photons are produced up to us by taking $g_{a\gamma\gamma}=0.5 \times 10^{-11} \, \rm GeV^{-1}$ and $m_a \lesssim 10^{-14} \, \rm eV$. In the first row we assume the peculiar case in which the line of sight is exactly coincident with the jet axis, while in the second row the common situation of a misalignment between the jet axis and the line of sight is considered. In all the panels we take $n_{e,0}^{\rm clu} = 5 \times 10^{-2} \, \rm cm^{-3}$, $B_{\rm ext} = 1 \, \rm nG$ and a redshift $z=0.03$. The initial photon degree of linear polarization is $\Pi_{L,0}=0.3$ in the left panels and $\Pi_{L,0}=0$ in the right panels.}
\end{figure*}

In order to understand if this is what really occurs, we plot in Fig.~\ref{densProbBlazObserv} the probability density function $f_{\Pi}$ of $\Pi_L$ associated to several realizations of the photon-ALP beam propagation process by considering the limited spatial resolution of the polarimeters in different situations. We consider the case: $B_{\rm ext} = 1 \, \rm nG$, a redshift $z=0.03$ and $n_{e,0}^{\rm clu} = 5 \times 10^{-2} \, \rm cm^{-3}$, as usual for a CC galaxy cluster hosting a blazar. Other cases are totally similar. We study the behavior in the X-ray (at $E_0 = 1 \, \rm keV$) and HE (at $E_0 = 10 \, \rm MeV$) band in the left and right panels of Fig.~\ref{densProbBlazObserv}, respectively. In the first row we examine the peculiar case in which the line of sight coincides exactly with the jet axis, while in the second row we consider the general situation of a misalignment between the jet axis and the line of sight by assuming the intermediate value $\theta_{\rm com} = 3 \pi/10$, which corresponds to $\theta_{\rm fix} \simeq 1/(2\gamma)$. 

In the former case, i.e. of perfect alignment, we have photons with equal probability of coming from whatever position in jet transverse section (see case a) of Fig.~\ref{imageJet}) and we thus simply average the Stokes parameters without any weight about the averaging procedure: results are shown in the first row of Fig.~\ref{densProbBlazObserv}. In the latter case, i.e with misalignment, we observe a larger amount of photons coming from a particular region of the jet because of the observation geometry under which we see the jet (see also Fig.~\ref{imageJet} for comparison), while photons from all the rest of the jet are much less abundant: the present situation is explored in the second row of Fig.~\ref{densProbBlazObserv}, where we weight the Stokes parameters through a gaussian distribution centered on that particular angular position in the jet transverse section, which is identified by the above defined plane.

Therefore, in the case of perfect alignment -- see the first row of Fig.~\ref{densProbBlazObserv} -- the final $\Pi_L$ values are reduced both in the X-ray and in the HE range because of the {\it flat} averaging procedure. 
 Instead, in the case of misalignment -- see the second row of Fig.~\ref{densProbBlazObserv} -- the final $\Pi_L$ values are consistent with those reported in the figures of the previous Sections.

The reason for this different behavior is that, when the line of sight coincides exactly with the jet axis, the {\it flat} averaging procedure -- for photons with equal probability of arrival -- mildly washes out the ALP-induced polarization features, since the photon-ALP beam experiences ${\bf B}^{\rm jet}$ with orientations in whatever direction, as discussed above (see also Fig.~\ref{imageJet}). Instead, in the general case of a misalignment between the jet axis and the line of sight, 
the {\it weighted} averaging procedure almost completely preserves the ALP-induced polarization features, since the photon-ALP beam experiences an almost coherent ${\bf B}^{\rm jet}$, as explained above (see also Fig.~\ref{imageJet}).

We have to consider also that the case of perfect alignment never occurs in realistic situations and that the region of close alignment is statistically very unlikely. Instead, the condition of misalignment is the most common one simply for purely statistical reasons. As a result, the final $\Pi_L$ is only marginally affected at its highest values by the limited spatial resolution of the instruments. Therefore, what we report in the second row of Fig.~\ref{densProbBlazObserv} and in the figures of the previous Sections represents the most common situation.

We plan to deep the analysis of the previous topic in a dedicated paper about ALP-induced effects on the polarization of photons produced in the blazar jet.

\section{Conclusions}

In this paper, we have studied the propagation of the photon-ALP beam up to the Earth when photons are produced in the central region of a nCC galaxy cluster ($n_{e,0}^{\rm clu} = 0.5 \times 10^{-2} \, \rm cm^{-3}$) and when they are generated in the jet of a blazar by accordingly considering a hosting CC galaxy cluster ($n_{e,0}^{\rm clu} = 5 \times 10^{-2} \, \rm cm^{-3}$). We have analyzed all the magnetized media crossed by the beam: the blazar jet, the host galaxy, the galaxy cluster, the extragalactic space and the Milky Way. We have considered the case of both efficient ($B_{\rm ext} = 1 \, \rm nG$) and negligible ($B_{\rm ext} < 10^{-15} \, \rm G$) photon-ALP conversion in the extragalactic space. In the presence of the photon-ALP interaction, we have then calculated the photon survival probability $P_{\gamma \to \gamma}$ and the corresponding final photon degree of linear polarization $\Pi_L$ by taking physically consistent values for the parameters concerning both the crossed media (magnetic field, electron number density and their profiles) and the photon-ALP system with $g_{a\gamma\gamma}=0.5 \times 10^{-11} \, \rm GeV^{-1}$ and two cases concerning the ALP mass: (i) $m_a \lesssim 10^{-14} \, \rm eV$, (ii) $m_a = 10^{-10} \, \rm eV$ (see also~\cite{noteFabian}). We have considered three energy ranges: (i) UV-X-ray band ($10^{-3}\, {\rm keV} - 10^2 \, \rm keV$), (ii) HE band ($10^{-1}\, {\rm MeV}-10^4 \, \rm MeV$), (iii) VHE band ($10^{-2}\, {\rm TeV}-10^3 \, \rm TeV$). While our results about the first two energy ranges can be tested by current and planned observatories~\cite{ixpe,extp,xcalibur,ngxp,xpp,cosi,eastrogam1,eastrogam2,amego}, our findings in the VHE band are currently of theoretical nature. We have checked that our results about $P_{\gamma \to \gamma}$ and $\Pi_L$ satisfy the theorems linking the conversion/survival probability and the initial photon polarization, which have been enunciated and demonstrated in~\cite{galantiTheorems}. Our results can be summarized as follows.

\begin{enumerate}[(i)]

\item In the UV-X-ray band, we take an initial photon degree of linear polarization $\Pi_{L,0}=0$ for the case of photon production in the cluster and $\Pi_{L,0}=0.3$ for the case of photon generation in the blazar jet, as explained in Sec. V.A. If $m_a = 10^{-10} \, \rm eV$, the photon-ALP conversion is very inefficient so that $P_{\gamma \to a} \to 0$ and ALP-induced effects on the final photon polarization are negligible. If $m_a \lesssim 10^{-14} \, \rm eV$, the photon-ALP beam propagates in the weak mixing regime for almost all the energy interval: $P_{\gamma \to \gamma}$ and the corresponding final $\Pi_L$ show oscillations with respect to the observed energy $E_0$. The probability density function $f_{\Pi}$ of $\Pi_L$ associated to several realizations of the photon-ALP beam propagation process shows that in all considered cases $\Pi_L$ is modified, broadened and its most probable expectation translates to a higher value with respect to the initial $\Pi_{L,0}$. 

\item In the HE band, we consider $\Pi_{L,0}=0$ for both the cases of photon production in the cluster and in the blazar jet, as explained in Sec. V.B. If $m_a \lesssim 10^{-14} \, \rm eV$, the photon-ALP beam propagates in the strong mixing regime in this energy interval, so that $P_{\gamma \to \gamma}$ and $\Pi_L$ are energy independent. In all considered cases, $f_{\Pi}$ shows a modification and broadening of the values assumed by the final $\Pi_L$ with respect to the initial $\Pi_{L,0}$. The most probable expectation for the final $\Pi_L$ is $\Pi_L \gtrsim 0.8$ but with a wide broadening. This fact can be understood because of the efficiency of the photon-ALP conversion that occurs in the strong mixing regime. Instead, if $m_a = 10^{-10} \, \rm eV$, the photon-ALP system lies in the weak mixing regime with a resulting oscillatory behavior with respect to the observed energy $E_0$ of both $P_{\gamma \to \gamma}$ and $\Pi_L$. The probability density function $f_{\Pi}$ of $\Pi_L$ shows that, in the case $m_a = 10^{-10} \, \rm eV$, $\Pi_L$ is less modified when photons are generated in the cluster.

\item In the VHE band, we take $\Pi_{L,0}=0$ for both the cases of photon production in the cluster and in the blazar jet, as discussed in Sec. V.C. For almost all the considered energy interval, the photon-ALP beam propagates in the weak mixing regime for both the cases $m_a \lesssim 10^{-14} \, \rm eV$ and $m_a = 10^{-10} \, \rm eV$, so that $P_{\gamma \to \gamma}$ and $\Pi_L$ show oscillations with respect to $E_0$. In addition, $f_{\Pi}$ still shows a modification and broadening of the values assumed by the final $\Pi_L$ with respect to the initial $\Pi_{L,0}$ but with a difference with respect to the previous energy intervals. In the VHE band $\gamma \gamma$ absorption caused by the EBL decreases the amount of photons which can be detected at the Earth. Therefore, we find a peculiar feature: when absorption is very high, all photons are absorbed in the extragalactic space, so that only photons reconverted back from ALPs in the Milky Way can be detected. In this case, the corresponding final $\Pi_L$ increases to very high values up to the limit $\Pi_L=1$ with almost no broadening, as shown by $f_{\Pi}$. Thus, a detection of fully polarized photons would represent a {\it proof} for the existence of ALPs with the properties discussed in this paper. However, the possibility of such a detection is nowadays only a hope for the future, since current techniques to measure photon polarization reach a few tens of GeV at most~\cite{polLimit}.

\end{enumerate}


We want to stress that we have assumed physically consistent parameters about the media crossed by the photon-ALP beam. By considering different values and profiles concerning the magnetic fields and the electron number densities (e.g. in the galaxy cluster), all our findings still hold true but with a translation to lower/higher energies of the weak mixing regime. 

As discussed above, observatories in the X-ray and HE bands~\cite{ixpe,extp,xcalibur,ngxp,xpp,cosi,eastrogam1,eastrogam2,amego} are expected to possess a sufficient energy resolution to be able to detect the photon polarization features induced by ALPs and analyzed in this paper. Still, we plan to study the actual detectability in further publications.

When photon polarization accurate data will be available, their analysis will be crucial, in order to understand their physical origin and to distinguish among several possibilities. In particular, since photons originated in the central region of a galaxy cluster are expected to be unpolarized both in the X-ray and HE bands, a detection of $\Pi_L > 0$ would represent a hint for new physics. Although Lorentz invariance violation (LIV) induces a variation to the final $\Pi_L$, LIV has the tendency of reducing $\Pi_L$~\cite{LIVpol}, so that a detection of $\Pi_L > 0$ for photons produced in the central zone of a galaxy cluster would invariably represent a hint for the existence of an ALP. In the case of photon generation in the blazar jet, the situation is more involved: for photons in the HE range everything we have just stated above still holds true since $\Pi_{L,0}=0$ in leptonic emission models~\cite{footnoteBla}. Instead, in the X-ray band, since $\Pi_{L,0}=0.2 - 0.4$, a final $\Pi_L \lesssim 0.1 - 0.2$ would represent a hint for LIV or ALPs, while a detected $\Pi_L \gtrsim 0.4 - 0.5$ would imply an indication for the existence of an ALP.


Finally, ALPs with the properties considered in this paper can be observed by the new generation of gamma-ray observatories such as CTA~\cite{cta}, HAWC~\cite{hawc}, GAMMA-400~\cite{g400}, LHAASO~\cite{lhaaso}, TAIGA-HiSCORE~\cite{desy} and HERD~\cite{herd}. Moreover, these ALPs can be directly detected by laboratory experiments like the upgrade of ALPS II at DESY~\cite{alps2}, the planned IAXO~\cite{iaxo,iaxo2} and STAX~\cite{stax}, and with other techniques developed by Avignone and collaborators~\cite{avignone1,avignone2,avignone3}. In addition, if ALPs are the greatest constituents of the dark matter, they can also be detected by the planned ABRACADABRA experiment~\cite{abracadabra}.

\section*{Acknowledgments}

The author thanks Enrico Costa, Marco Roncadelli and Fabrizio Tavecchio for discussions.
The work of the author is supported by a contribution from the grant ASI-INAF 2015-023-R.1.


\begin{thebibliography}{20}
\bibitem{foot1} Other processes discussed in~\cite{gptr} are totally irrelevant for the energy ranges considered in this paper.
\bibitem{noteClu} Other models with a larger number of parameters with respect to those entering Eq.~(\ref{eq2}) are employed to describe $n_e^{\rm clu}$ especially concerning CC clusters (see e.g.~\cite{betaModel,cluValues}). However, we have checked that the final results about $P_{\gamma \to \gamma}$ and $\Pi_L$ in the presence of the photon-ALP interaction are not substantially affected by the utilized model. Thus, we always employ Eq.~(\ref{eq2}) to describe $n_e^{\rm clu}$, in order to reduce the number of the system parameters.
\bibitem{noteFabian} Concerning the photon-ALP system parameters ($m_a, g_{a\gamma\gamma}$), although the only firm bound is that derived by CAST~\cite{cast}, there are indications that the model ($m_a=10^{-10} \, {\rm eV},g_{a\gamma\gamma}=0.5 \times 10^{-11} \, {\rm GeV^{-1}}$) is preferable with respect to that ($m_a \lesssim 10^{-14} \, {\rm eV}, g_{a\gamma\gamma}=0.5 \times 10^{-11} \, {\rm GeV^{-1}}$)~\cite{limFabian,limJulia,limKripp,limRey2}. Yet, the model ($m_a \lesssim 10^{-14} \, {\rm eV}, g_{a\gamma\gamma}=0.5 \times 10^{-11} \, {\rm GeV^{-1}}$) cannot be excluded.
\end{thebibliography}
\end{document}